\documentclass[sigconf,screen]{acmart}

\usepackage[utf8]{inputenc}

\usepackage{color}

\definecolor{grayblue}{rgb}{0.2,0.29,0.79}
\definecolor{darkgreen}{rgb}{0.2,0.55,0.1}
\definecolor{violet}{rgb}{0.54,0.17,0.88}

\definecolor{royalblue}{RGB}{65,105,225}

\newcommand{\xiao}[1]
{
   {\noindent\color{pink}\bf [#1]$_{\scriptscriptstyle\textit{xiao}}$}
}
\newcommand{\guo}[1]
{
   {\noindent\color{grayblue}\bf [#1]$_{\scriptscriptstyle\textit{guo}}$}
}
\newcommand{\gao}[1]
{
   {\noindent\color{violet}\bf [#1]$_{\scriptscriptstyle\textit{gao}}$}
}

\newcommand{\fang}[1]
{
   {\noindent\color{red}\bf [#1]$_{\scriptscriptstyle\textit{fang}}$}
}

\newcommand{\includeAuthorComments}[1]
{
   \ifthenelse{\equal{#1}{0}}
   {
      \renewcommand{\xiao}[1]
      {
         {} 
      }
      \renewcommand{\guo}[1]
      {
         {} 
      }
      \renewcommand{\gao}[1]
      {
         {} 
      }
      \renewcommand{\fang}[1]
      {
         {} 
      }
      \renewcommand{\chen}[1]
      {
         {} 
      }
   }{}
}

\AtBeginDocument{%
  \providecommand\BibTeX{{%
    \normalfont B\kern-0.5em{\scshape i\kern-0.25em b}\kern-0.8em\TeX}}}

\graphicspath{{figs/}}
\usepackage{graphicx}
\usepackage{subfigure}
\usepackage{booktabs}
\usepackage{bigstrut,multirow,rotating,booktabs}
\usepackage{amsmath}
\usepackage{CJKutf8}
\usepackage{algorithm}
\usepackage{algorithmicx}
\usepackage{algpseudocode}
\usepackage{amsmath}
\usepackage{microtype}
\floatname{algorithm}{Algorithm}

\usepackage{hyperref}
\hypersetup{
    colorlinks=true,            
    linkcolor=blue,             
    filecolor=magenta,          
    urlcolor=cyan,              
    pdftitle={Overleaf Example},
    pdfpagemode=FullScreen,
    }

\setcopyright{acmlicensed}
\acmDOI{10.1145/3650212.3680373}
\acmYear{2024}
\copyrightyear{2024}
\acmISBN{979-8-4007-0612-7/24/09}
\acmConference[ISSTA '24]{Proceedings of the 33rd ACM SIGSOFT International Symposium on Software Testing and Analysis}{September 16--20, 2024}{Vienna, Austria}
\acmBooktitle{Proceedings of the 33rd ACM SIGSOFT International Symposium on Software Testing and Analysis (ISSTA '24), September 16--20, 2024, Vienna, Austria}
\acmSubmissionID{issta24main-p1200-p}
\received{2024-04-12}
\received[accepted]{2024-07-03}

\begin{document}

\title{CooTest: An Automated Testing Approach for V2X Communication Systems}

\author{An Guo}
\email{guoan218@smail.nju.edu.cn}
\affiliation{
  \institution{State Key Laboratory for Novel Software Technology \\Nanjing University}
  \city{Nanjing 210023}
  \postcode{210023}
\country{China}
}
\orcid{0009-0005-8661-6133}

\author{Xinyu Gao}
\email{xinyugao@smail.nju.edu.cn}
\affiliation{
  \institution{State Key Laboratory for Novel Software Technology \\Nanjing University}
  \city{Nanjing 210023}
  \postcode{210023}
\country{China}
}
\orcid{0009-0004-7135-1833}

\author{Zhenyu Chen}
\authornote{Zhenyu Chen and Chunrong Fang are the corresponding authors.}
\email{zychen@nju.edu.cn}
\affiliation{
  \institution{State Key Laboratory for Novel Software Technology \\Nanjing University}
  \city{Nanjing 210023}
  \postcode{210023}
\country{China}
}
\orcid{0000-0002-9592-7022}

\author{Yuan Xiao}
\email{yuan.xiao@smail.nju.edu.cn}
\affiliation{
  \institution{State Key Laboratory for Novel Software Technology \\Nanjing University}
  \city{Nanjing 210023}
  \postcode{210023}
\country{China}
}
\orcid{0009-0009-3166-8007}

\author{Jiakai Liu}
\email{liujk@smail.nju.edu.cn}
\affiliation{
  \institution{State Key Laboratory for Novel Software Technology \\Nanjing University}
  \city{Nanjing 210023}
  \postcode{210023}
\country{China}
}
\orcid{0009-0005-1651-0720}

\author{Xiuting Ge}
\email{dg20320002@smail.nju.edu.cn}
\affiliation{
  \institution{State Key Laboratory for Novel Software Technology \\Nanjing University}
  \city{Nanjing 210023}
  \postcode{210023}
\country{China}
}
\orcid{0000-0003-3683-7374}

\author{Weisong Sun}
\email{weisongsun@smail.nju.edu.cn}
\affiliation{
  \institution{State Key Laboratory for Novel Software Technology \\Nanjing University}
  \city{Nanjing 210023}
  \postcode{210023}
\country{China}
}
\orcid{0000-0001-9236-8264}

\author{Chunrong Fang}
\authornotemark[1]
\email{fangchunrong@nju.edu.cn}
\affiliation{
  \institution{State Key Laboratory for Novel Software Technology \\Nanjing University}
  \city{Nanjing 210023}
  \postcode{210023}
\country{China}
}
\orcid{0000-0002-9930-7111}

\begin{abstract}

Perceiving the complex driving environment precisely is crucial to the safe operation of autonomous vehicles. With the tremendous advancement of deep learning and communication technology, Vehicle-to-Everything (V2X) collaboration has the potential to address limitations in sensing distant objects and occlusion for a single-agent perception system. However, despite spectacular progress, several communication challenges can undermine the effectiveness of multi-vehicle cooperative perception. The low interpretability of Deep Neural Networks (DNNs) and the high complexity of communication mechanisms make conventional testing techniques inapplicable for the cooperative perception of autonomous driving systems~(ADS). Besides, the existing testing techniques, depending on manual data collection and labeling, become time-consuming and prohibitively expensive.

In this paper, we design and implement CooTest, the first automated testing tool of the V2X-oriented cooperative perception module. CooTest devises the V2X-specific metamorphic relation and equips communication and weather transformation operators that can reflect the impact of the various cooperative driving factors to produce transformed scenes. 
Furthermore, we adopt a V2X-oriented guidance strategy for the transformed scene generation process and improve testing efficiency. We experiment CooTest with multiple cooperative perception models with different fusion schemes to evaluate its performance on different tasks. The experiment results show that CooTest can effectively detect erroneous behaviors under various V2X-oriented driving conditions. Also, the results confirm that CooTest can improve detection average precision and decrease misleading cooperation errors by retraining with the generated scenes.

\end{abstract}




\maketitle

\section{Introduction} 




Autonomous driving has attracted increasing attention due to its great potential to reduce the burden on drivers and improve traffic safety~\cite{yurtsever2020survey, liu2020computing}. As an exemplar of safety-critical intelligent software, autonomous driving systems rely on perception components to understand information about the surrounding environment and communicate perception results to downstream decision modules to enable smooth autonomous operation. Recent advancements in deep learning and sensor technology have improved the performance of modern perception systems~\cite{li2020lidar,wen2022deep}. Despite the remarkable progress, single-agent perception systems still yield inaccurate or incomplete perception results due to inherent single-view constraints (e.g., occlusion and distant sparse sensor observation), which can further lead to incorrect system behavior or even severe accidents~\cite{chen2022milestones, crash}. 
An illustration of this issue involves an Uber autonomous vehicle that collided with a pedestrian crossing the street.
The vehicle's perception system initially misidentified the pedestrian at a distance owing to single-view constraints. When the system eventually recognized the pedestrian, it was too late to avoid the collision~\cite{crash-bicycle1}.

Vehicle-to-Everything (V2X) communication has recently received significant attention as a solution to avoid the inherent limitations of single-agent sensing~\cite{cui2022cooperative,ma2024macp}. As a key technology in V2X autonomous driving, multi-agent cooperative perception strategies enable multiple connected agents to share complementary perceptual information with each other, thus providing more accurate and comprehensive perception~\cite{wang2020v2vnet,xu2022v2x}. Various prominent manufacturers and organizations, such as Ford and Baidu, are actively developing V2X platforms to enhance and ensure the performance of autonomous driving perception~\cite{apollo-v2x,ford-v2x}.
However, when deployed in real-world environments, such communication systems across multiple intelligent agents could encounter intricate communication challenges, such as lossy communication~\cite{nasralla2014subjective} and communication latency~\cite{wang2020v2vnet}. In practice, corrupted or misaligned shared perception information from cooperative vehicles could hardly assist the ego agent in understanding the surroundings or even misleading it into making incorrect decisions. Based on the above reasons, it is crucial to design a test approach for V2X cooperative perception systems.


Software engineering researchers have proposed several testing techniques to assess the potential risks of perception systems in real-world scenarios~\cite{guo2022lirtest,christian2023generating,wang2020metamorphic}. 
However, existing testing techniques only focus on testing the single-agent perception system rather than considering specific communication issues during the cooperation process. Moreover, practitioners are required to manually collect data obtained from connected vehicles during real-world road testing while covering various scenarios against diverse adverse communication conditions demands substantial resource consumption~\cite{liu2017bigroad}.
Furthermore, manual labeling for collected data (e.g., images, point clouds, etc.) from various connected vehicles is a laborious and time-consuming task requiring a massive workforce to manually examine the visualization results~\cite{kim2020reducing}.

To bridge this gap, we propose a systematic V2X-oriented automated testing approach and implement it into a tool
, namely CooTest. 
The primary objective of CooTest is to automate the testing of cooperative perception systems in self-driving cars and subsequently facilitate their performance improvement through retraining. 
CooTest is built upon the theory of metamorphic testing to reduce manual data collection and labeling efforts. It implements the metamorphic relation with two families of transformation operators that can reflect the characteristics of practical V2X cooperative perception scenes.
CooTest leverages a V2X-oriented guided approach in the testing process to efficiently generate testing cases derived from seed testing data utilizing transformation operators.
Furthermore, CooTest leverages transformation-specific metamorphic relations between the transformed data and seed data to detect the cooperative perception faults automatically.


We conduct experiments involving six state-of-the-art cooperative perception models with distinct information fusion schemes to assess the efficacy of CooTest. In the experiment, all transformation operators implemented by CooTest proved effective on models across various fusion schemes during the testing process. Subsequent experimental findings underscore that CooTest with V2X-oriented transformation guidance can generate more failed tests on cooperative perception models than random strategy guidance. Moreover, our experiments reveal that retraining the cooperative perception model with generated data can substantially decrease the misleading cooperation errors and enhance the performance by 6.7\% on average, as measured by average precision (AP).

The main contributions of this paper are summarized as follows:

\begin{itemize}

\item We propose an automated testing approach for cooperative perception systems based on the metamorphic testing theory. We implement metamorphic relations with two families of transformation operators designed to simulate influential factors in application scenarios, thereby generating test scenes. 
Under the V2X-oriented 
transformation guidance, CooTest can detect cooperative-related faults for cooperative perception systems and enhance them efficiently.

\item We implement the proposed approach into the automated testing tool, CooTest. To the best of our knowledge, CooTest is the first systematic and automated testing tool for V2X-oriented cooperative perception systems. To support the open science community, we have made the source code available$\footnote{https://github.com/meng2180/CooTest}$ and released the generated transformed data.

\item We utilize CooTest to assess six typical cooperative perception models and find various misleading cooperation errors, some of which may lead to potentially fatal collisions. The results show that the synthetic test cases can be used for retraining and making cooperative perception systems more robust to various corner cases.


\end{itemize}


\section{Cooperative Driving Systems}

\subsection{V2X in Autonomous Driving}

\begin{figure}[htbp]
	\centering
    \includegraphics[width=\linewidth]{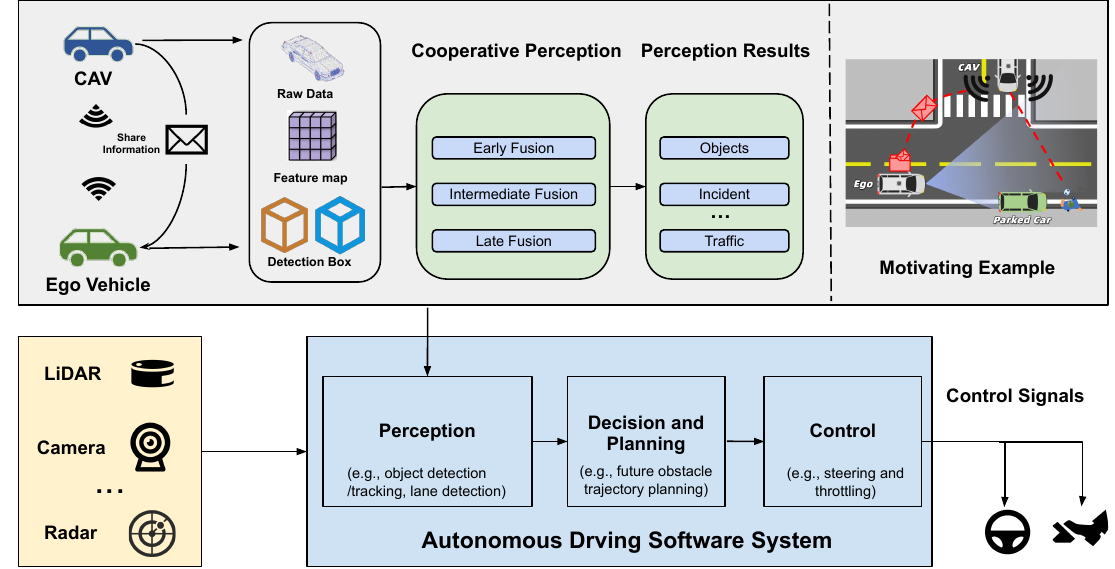}%
	\caption{The architecture of cooperative perception systems in autonomous driving. }
	\label{fig1}
    \vspace{-5pt}
\end{figure}

Autonomous driving systems is a complex system composed of various subsystems that operate sequentially.
Figure~\ref{fig1} shows the architecture of an autonomous driving software system based on modern designs like Autoware~\cite{ autoware} and Apollo~\cite{apollo}. 
The perception module is responsible for processing environmental information collected by various sensors to achieve functionalities such as object detection and tracking.
Upon receiving the processed perception information, the decision and planning module estimates future trajectories for the detected obstacles across both temporal and spatial dimensions. Finally, the control module calculates and executes the planned trajectory with lateral and longitudinal control.
The perception system forms the fundamental basis of autonomous driving, which is essential for understanding the environment and ensuring the smooth operation of the entire system.

Considering autonomous driving systems are operated in safety-critical contexts and perception systems play a crucial role, most industrial manufacturers leverage the V2X technique to overcome inherent single-view perception limitations in single-agent sensing (e.g., occlusions and short-range perceiving capability) and improve the overall performance~\cite{cui2022cooperative}. V2X denotes vehicles receiving shared information from other traffic elements via wireless communication technology~\cite{yu2021edge}.
The received information is rectified, matched, and integrated with ego vehicle sensing data into comprehensive and complementary perception information. 
This integrated data is then utilized by the perception system as input to make more accurate environmental predictions.
Such a process involving multi-agent collaborative perception lays the foundation of the V2X autonomous driving workflow and significantly impacts the final ADS behaviors, as shown in Figure~\ref{fig1}.
V2X cooperative autonomous driving has demonstrated outstanding performances in intelligent traffic systems, encompassing tasks such as road intersection management~\cite{chow2021adaptive} and the design of vehicle platooning~\cite{li2018eco}.

\subsection{V2X Cooperative Perception with On-road Agents}
\label{bg: cp}


Owing to the inherent constraints associated with the camera and LiDAR devices, the issues of occlusions and long-distance perception pose considerable challenges for autonomous systems designed for single vehicles~\cite{xu2022opv2v}. These challenges can give rise to severe consequences within intricate traffic scenarios. In contrast, cooperative perception systems can unlock the potential for multi-vehicle detection, effectively addressing the constraints associated with single-vehicle perception. By leveraging the V2X communication technology, different connected vehicles and infrastructures can share their sensing information and thus provide multiple viewpoints for the same obstacle to compensate each other~\cite{hobert2015enhancements,jung2020v2x}.


V2X cooperative perception methods can be divided into three categories based on the information fusion stage: \textit{early fusion}, \textit{late fusion}, and \textit{intermediate fusion}~\cite{han2023collaborative}. Early fusion techniques directly share raw data with connected agents located within the communication range. Then, the ego vehicle utilizes the aggregated data to predict detected objects~\cite{2019-Cooper}. In contrast, late fusion methods transmit detection outputs and integrate received proposals to formulate a coherent prediction~\cite{rawashdeh2018collaborative}. 
To meet the bandwidth and detection accuracy requirements, researchers have delved into intermediate fusion~\cite{2019-F-cooper, wang2020v2vnet}.
Intermediate fusion encodes the sparse raw data into dense vector representations for sharing.

V2X cooperative perception system holds the potential to enhance the safety and reliability of autonomous vehicles through the exchange of information with surrounding vehicles. However, it may face challenges from environmental factors like adverse weather (e.g., rain, snow, fog, etc.). Additionally, cooperative perception may introduce communication delays and unavoidable errors in relative pose among vehicles during data transmission. Therefore, it is critical to ensure the quality of the V2X communication systems before deploying them in a real-world scenario.




\section{Approach}



\begin{figure*}[htbp]
	\centering
    \includegraphics[width=\linewidth]{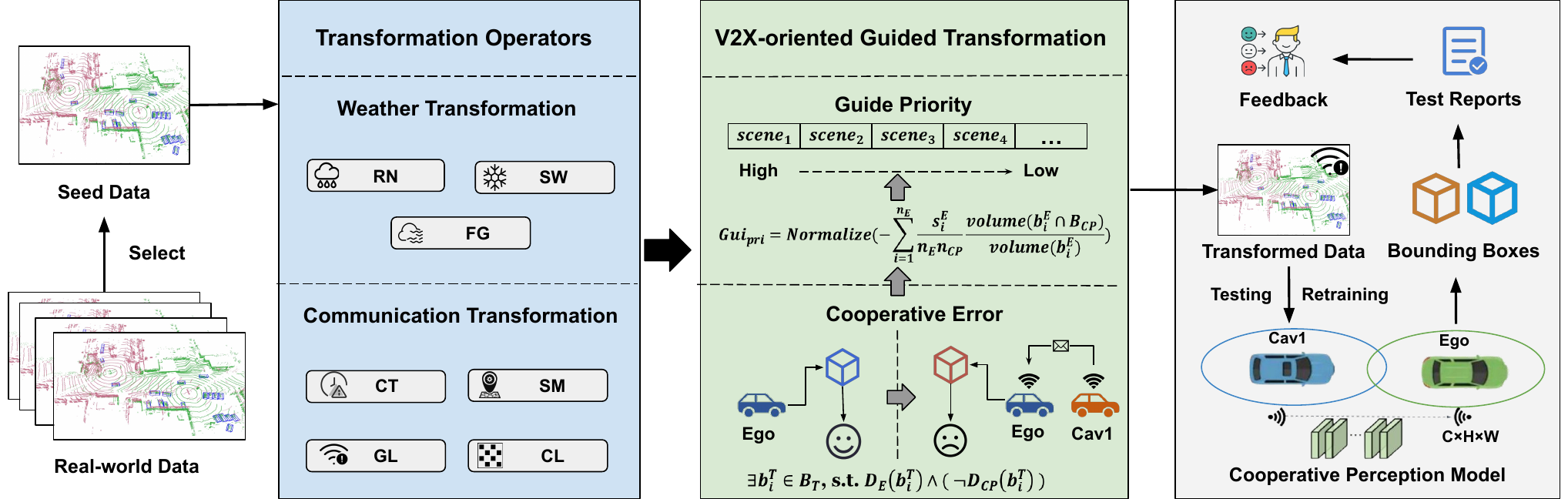}%
	\caption{The Workflow of CooTest.}
	\label{fig-structure}
\vspace{-5pt}
\end{figure*}

In this section, we present the design and implementation of CooTest, a tool devised for automatically testing the V2X-oriented cooperative perception module within ADS. In Figure\ref{fig-structure}, the developer initiates the process by selecting seeds from real-world data. Subsequently, CooTest adopts the metamorphic relation that we defined specifically for the cooperative perception systems. CooTest can generate new test scenes based on the seed by applying communication and weather transformation to simulate practical V2X application scenarios. Guided by V2X-oriented guidance, CooTest efficiently selects test data most likely to expose potential defects in the model. The transformed test scenes are then generated to assess the robustness of the cooperative perception model in autonomous driving. Finally, CooTest generates test reports containing AP and misleading cooperation errors based on models' predicted bounding boxes and gives feedback on model perception robustness. Moreover, CooTest can enhance the model's cooperative perception performance through retraining with generated transformed scenes.





\subsection{Formalization and Definition}\label{section:definition}

Cooperative errors in complex cooperative perception systems can arise from multiple components, including the ego vehicle, cooperative vehicles or infrastructures, and the information fusion process. The most basic reliability requirement of V2X systems is that the perception performance of the ego vehicle should not be impaired by the cooperative perception system, especially for close-range object detection~\cite{DBLP:journals/corr/abs-2401-01544}. However, due to communication challenges (Section~\ref{bg: cp}), shared information from cooperative vehicles may provide limited assistance to the ego agent in understanding surroundings, and in certain situations, it might even share misleading information. 
In this case, even if the sensors equipped on the ego vehicle provide correct environment understanding, the ego vehicle's perception system will still prioritize the inaccurate shared information from collaborators over the correct one.
In this paper,
we refer to this error due to incorrect cooperative perception as \textit{misleading cooperation error}. Before giving a formal definition of misleading cooperation error, we give a formal representation of V2X-oriented cooperation perception.



\subsubsection{Formal Representation}

We first formalize the V2X-oriented  cooperative perception system and the ego vehicle perception system.
Consider $n$ cooperative vehicles and an ego vehicle in the scene. Let $X_E$ and $X_i$ represent the observation of the ego vehicle and the $i$-th connected vehicle, respectively. The cooperative perception systems $\mathbb{CP}$ can be represented as: 
$$\Phi_{CP}(X_E,\{X_j\}_{j=1}^n)=\{(s_i^{CP},b_i^{CP})|i\in [n_{CP}]\}$$
Similarly, the ego vehicle perception system can be represented as:
$$\Phi_E(X_E)=\{(s_i^E,b_i^E)|i\in [n_E]\}$$
where $x\in \{E,CP\}$ and $\Phi_x$ is the perception network of perception system.  $b_i^x,s^x_i$ represent the $i$-th 3D bounding box and its confidence of the perception results, respectively. $[n_x]=1, 2, \cdots, n_x$ and $n_x$ represent the number of 3D bounding boxes detected by the perception system.  We leverage $S_x=\cup_{i=1}^{n_x} s^x_i, x\in\{E,CP\}$ representing the set of all $s^x_i$.  







\subsubsection{Misleading Cooperation Error~(MCE)}
Misleading cooperation error occurs (1) when an ego vehicle accurately perceives its surroundings during operation, and (2) the collaborative perception system is misled by the inaccurate shared information to make incorrect perception predictions.
Here, we define the misleading cooperation error of a cooperative perception system $\mathbb{CP}$. 
Let $b^T_i$ represent the $i$-th ground-truth bounding box and $n_T$ represent the number of ground-truth 3D bounding boxes. We use $B_x=\cup_{i=1}^{n_x} b^x_i, x\in\{T,E,CP\}$ representing the set of all $b^x_i$. A correct cooperative perception method should detect the bounding box, which is ground truth and successfully detected by the ego vehicle. Thus, a misleading cooperation error occurs if $\mathbb{CP}$ cannot detect the ground-truth bounding box, which is successfully detected by the ego vehicle.



\noindent \textbf{Definition.} Consider a cooperative perception system $\mathbb{CP}$.  $\mathbb{CP}$ makes misleading cooperation error if
\begin{equation}
\begin{aligned}
      \exists b_{i}^T\in B_T, 
s.t.~D_E(b_{i}^T)\wedge (\neg D_{CP}(b_{i}^T))
\end{aligned}
\nonumber
\end{equation}
$D_E$ and $D_{CP}$ are the criteria to decide whether the ego vehicle system and the cooperative perception system $\mathbb{CP}$ successfully detect the bounding box, respectively. 
Such misleading cooperation errors can diminish the precision of the cooperative perception system, thereby influencing the safety and reliability of autonomous driving. 
To this end, this paper focuses on misleading cooperation error and their impact on the overall perception performance of cooperative systems.

\subsection{Metamorphic Testing for Cooperative Perception System}\label{section:MR}

A significant challenge in testing the intricate cooperative perception system lies in manually creating the system's specifications. Moreover, annotating LiDAR-captured point clouds requires a significant amount of time.
Cooperative perception entails all connected autonomous vehicles participating in data collection, thus demanding a greater labeling effort compared to single-agent sensing. 
To overcome these issues, we leverage metamorphic relations (MRs)~\cite{chen2018metamorphic, 2020-Metamorphic-Testing} between the cooperative perception results across the original test data and its transformed test data to create the test oracle. The central idea is the ability to establish relationships among the model's detection outcomes under specific transformations. The violation of MRs often signals potential defects.

CooTest is designed to realize specific MRs designed for V2X communication systems. 
Specifically, given a seed test suite $\mathbb{T}_{orig}$ collected from connected vehicles, and a transformation operator $\gamma \in \mathbb{O}$ which can generate various transformed data from each original data $S \in \mathbb{T}_{orig}$, the MR to test cooperative perception system $\mathbb{CP}$ with additional transformed point clouds adopted in this paper can be formalized as follows:
\begin{equation}
\forall S \in \mathbb{T}_{orig} \wedge \forall \gamma \in \mathbb{O},  \zeta \{ \mathbb{C P} \llbracket \gamma(S) \rrbracket , \mathbb{C P} \llbracket S \rrbracket \}
\end{equation}

where $\zeta$ is a criterion asserting the equality of $\mathbb{CP}$ results. The given MR is defined such that no matter how the transformed data is synthesized by applying specifically designed transformation operators denoted as $\gamma$ on seed data $S$, the resulting object detection outcomes are anticipated to maintain consistency with those from the original data. 
Thus, if a $\mathbb{CP}$ infers a set of 3D bounding boxes $B_{o}$ for a seed data $S$ and infers another set of boxes $B_{c}$ for a corresponding transformed data $S_{c}$ generated by applying transformation $\gamma$ on $S$, $B_{0}$ and $B_{c}$ should be identical; otherwise, erroneous predictions can be revealed by checking the failure of this MR. 

However, asserting the strict equality of outputs from $\mathbb{CP}$ is often too stringent because there is usually no single correct detection result for a given test case, which may result in many false positives. 
To introduce more permissiveness into the MRs, the equality criteria $\zeta$ is derived from AP~\cite{everingham2010pascal}. The AP score is computed by considering both precision and recall values, as explained in Section~\ref{section:metric}. By utilizing V2X-oriented MRs, CooTest can straightforwardly validate their satisfaction within the testing process of $\mathbb{CP}$ by verifying against the test oracle. Based on this, CooTest further identifies misleading cooperative errors by automatically verifying the accuracy of the ego vehicle's perception results
In this context, the design and implementation of transformation operators become critical for optimizing the performance of CooTest.

\subsection{Transformation Operators}\label{section:transformation operators}

V2X communication systems operating in the real world frequently face unexpected environmental conditions. These conditions could give rise to subtle variations in the signals during the communication process, causing the ego vehicle to receive low-quality shared information and potentially resulting in unexpected behaviors.
Therefore, V2X communication systems are required to be robust against different corruption signals in various conditions.

To simulate practical application scenarios as much as possible, we investigate the common challenge for different V2X communication systems and design seven different realistic V2X-oriented transformation operators: communication latency (CT), spatial misalignment (SM), global feature lossy communication (GL), channel-specific lossy communication (CL), rain (RN), snow (SW), and fog (FG). These transformations can be classified into two categories: communication transformation and weather transformation. 
We implement these transformation operators and integrate them into CooTest. The detailed definition and design are as follows:




\subsubsection{Communication Transformation}


Due to the nature of the connectivity in the context of V2X, factors such as channel errors, network congestion, and delay violations can contribute to communication issues during data transmission within the wireless network~\cite{nasralla2014subjective}.
These communication issues can be divided into three categories, including \textit{communication latency}, \textit{spatial misalignment}, and \textit{lossy communication}~\cite{liu2023towards}.
Therefore, we design the corresponding communication transformation operators for each category and leverage them to simulate different real-world V2X-oriented communication issues. The specific operators are designed as follows:

\noindent\textbf{Communication Latency.} 
V2X communication demands well-synchronized data transmission across connected vehicles to ensure the quality and timeliness of shared sensory information.
In practical scenarios, limited communications bandwidth~\cite{wang2020v2vnet} or transmission failure~\cite{xu2022v2x} may cause a time delay in a participant, resulting in communication latency. 
We simulate the communication delay during the data transmission process, which can be expressed as:
\begin{equation}
\Delta M_{ego}(q)=M_{ego}^i -M_{cav_i}
\end{equation}
where $M_{ego}^i$ and $M_{cav_i}$ respectively represent the timestamp when the ego vehicle receives the data packet $q$ sent by the $i$-th connected vehicle and the corresponding data packet $q$ collected by the $i$-th connected vehicle, respectively. $\Delta M_{ego}$ can represent a time delay in transmitting data from the $i$-th connected autonomous vehicle to the ego vehicle.

\noindent\textbf{Spatial Misalignment.}
V2X communication system requires the determination of a positional transformation matrix between participants to ensure that local perception results from multiple participants can be matched efficiently.
However, obstacles or GPS signal interference~\cite{liu2021automated} in the real world will inevitably interfere with obtaining precise positioning information, leading to data misalignment during aggregation and performance degradation in cooperative perception. 
The transformation from the connected vehicle coordinates to the ego vehicle can be expressed as:
\begin{equation}
p_{\mathrm{cav}_{\text {projected }}}^t=T_{\text {cav } \rightarrow \mathrm{ego}} \cdot p_{\mathrm{cav}}^t
\end{equation}
where $p_{\text {cav }}^t$ is the pose $[x, y, z, 1]^{\mathrm{T}}$ in $i$-th connected autonomous vehicle at the time $t$, and $T_{\text {cav } \rightarrow \text { ego }} \in \mathbb{R}^{4 \times 4}$ is coordinate transformation matrix from connected vehicle to ego vehicle. 
In our experiments, we leverage rotation and translation to simulate spatial misalignment between the ego vehicle and connected vehicles.





\noindent\textbf{Lossy Communication.} Within urban traffic scenarios, various random factors, such as obstacle presence, rapid and fluctuating vehicle speeds, as well as varying distances between vehicles, may lead to lossy communication when transmitting a set of shared feature values~\cite{nasralla2014subjective,belyaev2014robust}. To simulate the realistic and complex lossy communication mechanisms in the real world, we designed two transformations based on the proposed communication models~\cite{li2023learning} to simulate different lossy types in real-world V2X communication. The specific details are as follows:

\begin{itemize}

\item \textbf{Global Feature Lossy Communication.} The shared feature after V2X metadata sharing is reshaped from a 3D tensor to a 2D matrix first. Subsequently, the reshaped feature undergoes random selection based on the global random probability $p_{g}$ and is replaced by random noise within the range of the original shared feature values.

\item \textbf{Channel-specific Lossy Communication.} Different from simulated lossy communication on the reshaped global feature, the channel-specific lossy type simulates lossy communication across distinct channels. Specifically, given a shared feature $C \times H \times W,\lfloor p_{c} * C\rfloor$ channels are randomly selected by the random probability $p_{c}$. Subsequently, these selected channels are replaced by random noise, maintaining values within the range of the original shared feature.

\end{itemize}

Eventually, the simulated lossy feature is transformed back to its initial shape of $C \times H \times W$ before being received by the ego vehicle.

\subsubsection{Weather Transformation}

Adverse weather conditions can induce measurement distortions in the sensors of each connected vehicle, presenting a substantial challenge for cooperative perception systems reliant on redundant information~\cite{heinzler2019weather}. For instance, droplets in the environment can decrease the detected object reflectivity and cause scattered points due to laser backscattering and absorption, leading to a reduction in the perceived quality of LiDAR point clouds. Therefore, we design the corresponding weather transformation operators to simulate different real-world weather.

In this study, we consider three representative adverse weather conditions: \textbf{Rain}, \textbf{Snow}, and \textbf{Fog}, as they are common in real-world driving scenarios.
To achieve this, we utilize domain-specific physical models as outlined in~\cite{2021-LISA} to simulate the characteristics of each weather condition. The realism of these models has been validated in several benchmark studies within the community~\cite{gao2023benchmarking,DBLP:journals/corr/abs-2210-05896}. During each transformation, we fix the environmental parameters and run the transformation for each connected vehicle to ensure that the connected vehicles are in an identical weather environment.




\begin{CJK*}{UTF8}{gkai}
    \begin{algorithm}
        \caption{V2X-oriented Guided Scene Transformation}
        \begin{algorithmic}[1] 
            \Require {The tested model $\mathbb{CP}$, the transformation operator set $\mathbb{A}$, the set of seed scene $\mathbb{T}_{orig}$, the number of generated tests $num_{gen}$}
            \Ensure  Generated high-quality tests $genTests$
            \\
            $genTests \leftarrow \emptyset$\;
            \For{$S$ in $\mathbb{T}_{orig}$}
                \For{$a$ in $\mathbb{A}$}
                \State $S^{'}=ApplyTransform(S,a)$
                    \State $S_E,n_E,n_{CP},B_E,B_{CP}=GetPred(\mathbb{CP},S^{'})$
                    \State $Gui_{pri}=CalculateGuiPri(S_E,n_E,n_{CP},B_E,B_{CP})$
                    \State $Spc=CreateTransScene(S^{'},Gui_{pri})$
                    \If{$len(genTests)<num_{gen}$}
                    \State $genTests.update(Spc)$
                    \Else
                        \If{$Spc.Gui_{pri}>genTests[-1].Gui_{pri}$}
                        \State $genTests[-1]=Spc$
                        \State $genTests.SortByGuiPri()$
                        \EndIf
                \EndIf
                \EndFor
            \EndFor
            \\
            \textbf{return} $genTests$
        \end{algorithmic}
        \label{alg1}
    \end{algorithm}
\end{CJK*}
\vspace{-10pt}

\subsection{V2X-oriented Guided Transformation}\label{section:guided method}
In theory, the number of tests that each transformation operator can generate is endless. Therefore, we must impose limitations and guidance to ensure the quality of tests and improve the testing efficiency.
To address this, 
inspired by the differential testing in traditional software testing, which leverages behavioral differences between multiple systems to detect errors ~\cite{mckeeman1998differential}, 
we propose a novel V2X-oriented guided transformation~(VGT) strategy designed for cooperative perception systems to bootstrap the test generation process. The basic idea behind this guidance strategy is to maximize the difference in environmental understanding between the ego vehicle perception and cooperative perception, thus generating test inputs that can confuse cooperative perception models, which can be formalized as follows:


\begin{equation}
    \label{eq: guide}
    \begin{aligned}
    Gui_{pri}&=Normalize(-\sum_{i=1}^{n_E} \frac{s_i^E}{n_E n_{CP}}\frac{volume(b^E_i\cap B_{CP})}{volume(b^E_i)})\end{aligned}
\end{equation}


$Gui_{pri}$ can be considered as a measure of the possibility of the cooperative perception system $\mathbb{CP}$ making misleading cooperation errors. The larger $Gui_{pri}$ means that the cooperative perception system $\mathbb{CP}$ is less capable of detecting the correct bounding box that the ego vehicle can detect. Specifically,
we compute the ratio of the intersection volume of the bounding box detected by both $\mathbb{CP}$ and the ego vehicle to the volume of the bounding box $b^E_i$ detected by the ego vehicle. We weigh the ratio by the confidence $s_i^E$ of the bounding box detected by the ego vehicle to emphasize the case that the ego vehicle detects the bounding box with high confidence while $\mathbb{CP}$ does not detect this bounding box. The above value is summed and then averaged by the number $n_E$ of the bounding boxes detected by the ego vehicle. We divide the value by the number $n_{CP}$ of bounding boxes detected by $\mathbb{CP}$ to reduce the effect that there exist two or multiple bounding boxes detected by $\mathbb{CP}$ overlap with one bounding box at the same time. Finally, we normalize the negative value of the result to quantify the rate of predicted misleading cooperation errors.
 As such, we get the resulting value $Gui_{pri}$.

This guidance strategy enables CooTest to uncover more model defects with the same testing set size, thereby improving testing efficiency. It involves sequentially traversing the seed scenes and randomly selecting operators for transformation, ensuring diversity in the algorithm's output.
Algorithm~\ref{alg1} presents the process of V2X scene transformation of CooTest in detail. The algorithm takes the tested model $\mathbb{CP}$, a set of communication transformation operators $\mathbb{A}$ with the corresponding parameters, a seed set $\mathbb{T}_{orig}$, and the number of generated tests $num_{gen}$ as input. Transformed tests can retain the same oracle information as the seed after a single transformation by configuring parameters for the transformation operators (Line 4). The core implementation process of this algorithm involves initially transforming the seed $S$ and subsequently calculating the guidance metric value based on the predictions of bounding boxes and confidence through Equation~\ref{eq: guide} (Lines 5-7).
Then, the generated test cases will be added to $genTests$ (Lines 8-9). When the generated quantity surpasses $num_{gen}$  (Lines 10-15), CooTest retains tests that are more likely to detect recognition errors.

\begin{figure*}[htbp]
	\centering
    \includegraphics[width=2\columnwidth,height=0.25\linewidth]{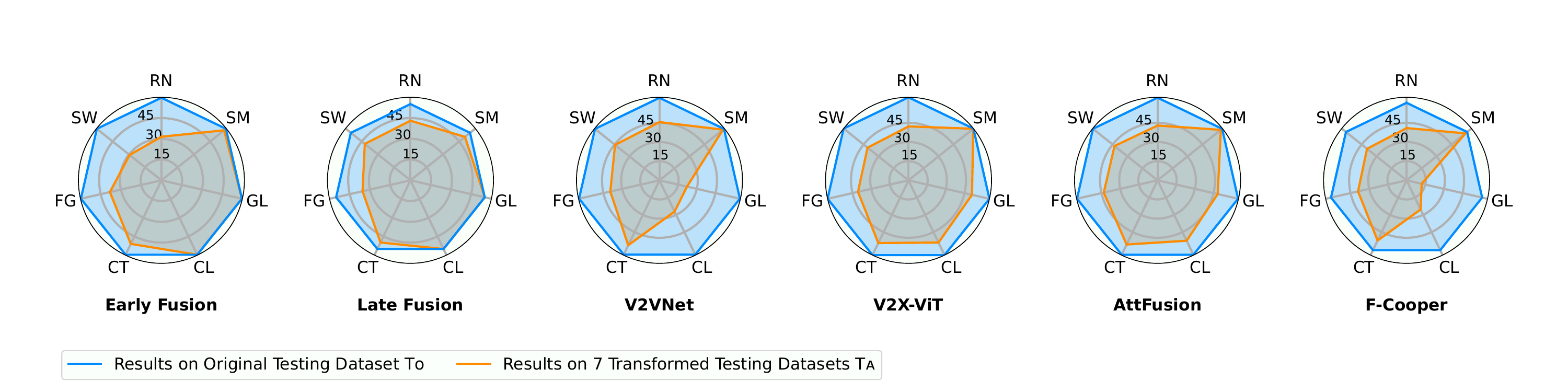}%
	\caption{Testing results on different models with seven transformation operators.}
	\label{fig:rq1}
 
\end{figure*}

\section{Experimental Design}
In this section, we introduce the experimental design, including the dataset and cooperative perception models under test, transformation settings, and evaluation metrics in the experiments. To conduct the experiments, we implemented the workflow of CooTest. All experiments are performed on Ubuntu 21.10 desktop with GeForce RTX 4070, one 16-core processor at 3.80GHz, and 32GB RAM.

CooTest is designed to systematically test the cooperation driving system, especially to verify its robustness for the V2X cooperative perception module. To this end, we empirically explore the following three research questions (RQ):
\begin{itemize}
\item RQ1: How effective are different transformation operators for detecting the erroneous behaviors of cooperative perception tasks using MRs?
\item RQ2: How effectively can CooTest generate tests under the V2X-oriented guided transformation?
\item RQ3: Can CooTest improve the cooperative perception model through retraining with transformed tests?


\end{itemize}



\subsection{Dataset and Cooperative Perception Models}

\textbf{V2V4Real}~\cite{xu2023v2v4real} is a large-scale real-world dataset for Vehicle-to-Vehicle cooperative perception. The data is collected by cooperative vehicles equipped with sensors, navigating through various scenarios. It comprises driving logs over three days, encompassing a total of 347 km on highways and 63 km within city roads. Additionally, it consists of 20,000 LiDAR frames, 40,000 RGB frames, 240,000 annotated 3D bounding boxes covering five classes, and comprehensive HDMaps encompassing all the driving routes. In this paper, we apply the proposed technique to the original test split of the cooperative object detection dataset to detect errors related to the V2X cooperative perception module.

\textbf{Cooperative Perception Models.} To assess the performance of our techniques comprehensively, we employ six cooperative perception models with diverse fusion schemes in our experiment, including one early fusion model, one late fusion model, and four state-of-the-art intermediate fusion models. All of these collaborative perception models are LiDAR-based and take multi-view point clouds as input. A brief introduction is provided as follows.

\textbf{\emph{Early Fusion}} model~\cite{xu2023v2v4real} directly transmits the raw point clouds to other collaborators, and the ego vehicle will aggregate all the data to its own coordinate frame. This process ensures the preservation of complete information.

\textbf{\emph{Late Fusion}} model~\cite{xu2023v2v4real} detects objects using the sensor observations of connected vehicles and subsequently shares the detection results with others. The receiving ego vehicle then applies non-maximum suppression to generate the final outputs.



\textbf{\emph{V2VNet}} model~\cite{wang2020v2vnet} proposes multi-round message passing based on graph neural networks to achieve better perception performance. It consists of three main stages: a convolutional network block to produce an intermediate representation, a cross-vehicle aggregation stage, and an output network to compute the final outputs.

\textbf{\emph{V2X-ViT}} model~\cite{xu2022v2x} introduces a novel vision transformer designed specifically for V2X perception. It incorporates a customized heterogeneous multi-head self-attention module crafted for graph attribute-aware multi-agent 3D visual feature fusion. This module is adept at capturing the inherent heterogeneity within V2X systems.



\textbf{\emph{F-Cooper}} model~\cite{2019-F-cooper} employes the maxout fusion to aggregate shared intermediate features. Data inputs undergo processing by the voxel feature encoding layers independently to generate voxel features. Next, spatial features are obtained locally on individual vehicles and fused together to generate the ultimate feature maps.


\textbf{\emph{AttFusion}} model~\cite{xu2022opv2v} is designed to capture interactions among features of neighboring connected vehicles, enabling the network to focus on key observations. This pipeline is flexible and can be seamlessly integrated with current deep learning-based detectors.


\begin{table}[htbp]\scriptsize
  \centering
  \caption{Transformation operators and parameters used by CooTest for generating new V2X-oriented scenes.}
    \begin{tabular}{c|c|c|c}
    \hline
    \textbf{Types} & \textbf{Operators} & \textbf{Parameters} & \textbf{Parameter Ranges} \bigstrut\\
    \hline
    \multirow{3}[6]{*}{\textbf{Weather}} & RN    & $r_{n}$    & (0.1,10) \bigstrut\\
\cline{2-4}          & SW    & $s_{w}$    & (0.1,2.4) \bigstrut\\
\cline{2-4}          & FG    & $f_{g}$    & (200,1000) \bigstrut\\
    \hline
    \multirow{5}[8]{*}{\textbf{Communication}} & CT    & $c_{t}$    & (0,300) \bigstrut\\
\cline{2-4}          & CL    & $p_{c}$    & (0,1) \bigstrut\\
\cline{2-4}          & GL    & $p_{g}$    & (0,1) \bigstrut\\
\cline{2-4}          & \multirow{2}[2]{*}{SM} & \multirow{2}[2]{*}{$s_{m}=(t_{x},r_{y},t_{z},r_{z})$} & $t_{x}\in$ (-0.2,0.2), $t_{y}\in$ (-0.2,0.2), \bigstrut[t]\\
          &       &       & $t_{z}\in$ (-0.2,0.2), $r_{z}\in$ (-2,2), \bigstrut[b]\\
    \hline
    \end{tabular}%
  \label{tab:tabpara}%
\end{table}%

\subsection{Transformations Settings} \label{section:4.2}
We leverage CooTest to improve the robustness of the cooperative detection model in adverse V2X-oriented driving conditions by making realistic communication and weather transformations. As described in Section~\ref{section:transformation operators}, we mainly apply seven different realistic data transformation operators: i.e., CT, SM, GL, CL, RN, SW, and FG. In order to make our transformations realistic, each transformation is designed to represent corresponding cooperative driving semantics in real-world scenarios. The RN, SW, and FG operators use rainfall intensity~(mm/h), snowfall intensity~(mm/h), and fog visibility~(m) as weather transformation semantics, respectively. The CL operator employs milliseconds (ms) as the unit for the transmission delay in vehicle-to-vehicle communication. For the SM operator, CooTest translates a specific distance (m) along the axis and rotates a certain angle around the z-axis. For lossy communication transformations, we leverage the global random probability $p_{g}$ and the probability of choosing a channel $p_{c}$ to simulate various lossy communication situations when transmitting a set of shared feature values. In our experiment design, candidate parameters for each transformation are shown in Table \ref{tab:tabpara}. Whenever the transformation operator is applied to the original data, CooTest randomly selects the parameters for a specific transformation.

\subsection{Evaluation Metric}\label{section:metric}

The primary task of cooperative perception is to collaboratively detect objects across multi-view 3D point clouds collected by the ego vehicle and the connected vehicles.
Identified objects are annotated with a 3D bounding box.
The central concept for evaluating the 3D object detection performance is Intersection over Union (IoU), which compares the coordinates of ground truth and predicted 3D bounding boxes~\cite{padilla2020survey}. The calculation of IOU can be represented as:

\begin{equation}
I o U=\frac{\operatorname{area}\left(B_{p} \cap B_{g t}\right)}{\operatorname{area}\left(B_{p} \cup B_{g t}\right)}
\end{equation}

Specifically, IoU measures the overlap (intersection) area between a ground-truth 3D bounding box in bird's-eye view $B_{gt}$ and a predicted 3D bounding box $B_p$ in bird's-eye view over their union. In our experiments evaluating cars, we adhere to the conventions established in other 3D object detection research~\cite{fernandes2021point}, requiring a bounding box overlap of at least 50\%.

\begin{figure*}[htbp]
    \centering
    \subfigure[No Collaboration]{
    \includegraphics[width=1.6in]{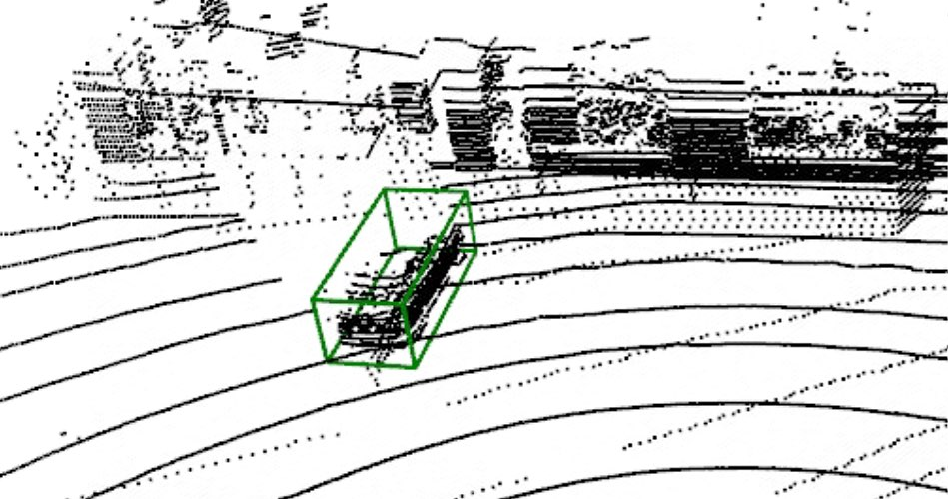}
    }
    \subfigure[Collaboration]{
	\includegraphics[width=1.6in]{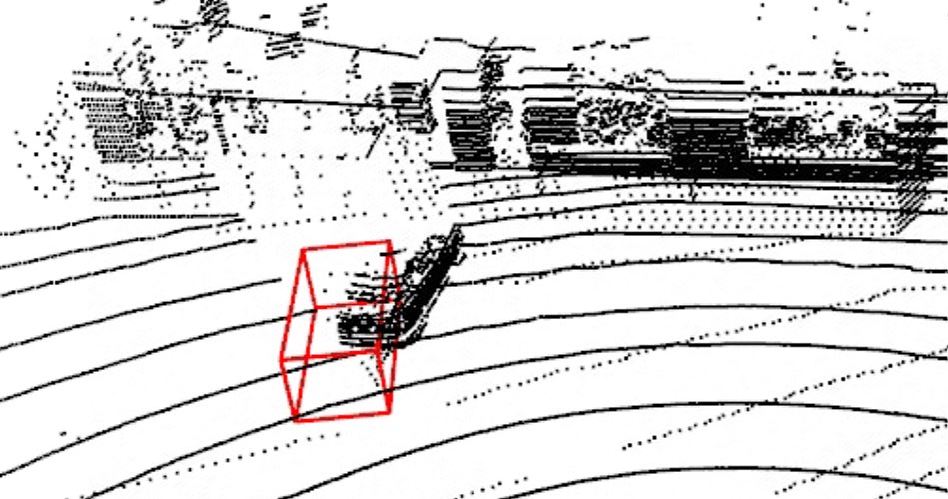}
    }
    \subfigure[No Collaboration]{
	\includegraphics[width=1.6in]{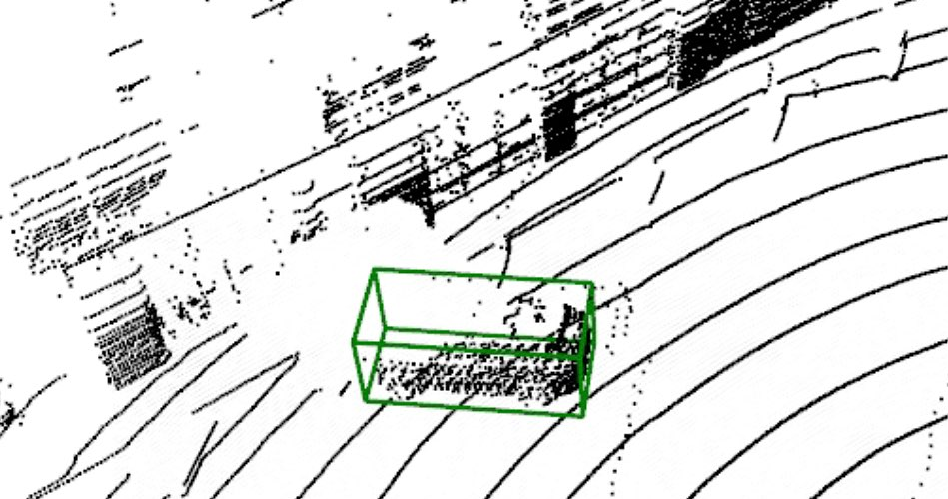}
    }
    \subfigure[Collaboration]{
	\includegraphics[width=1.6in]{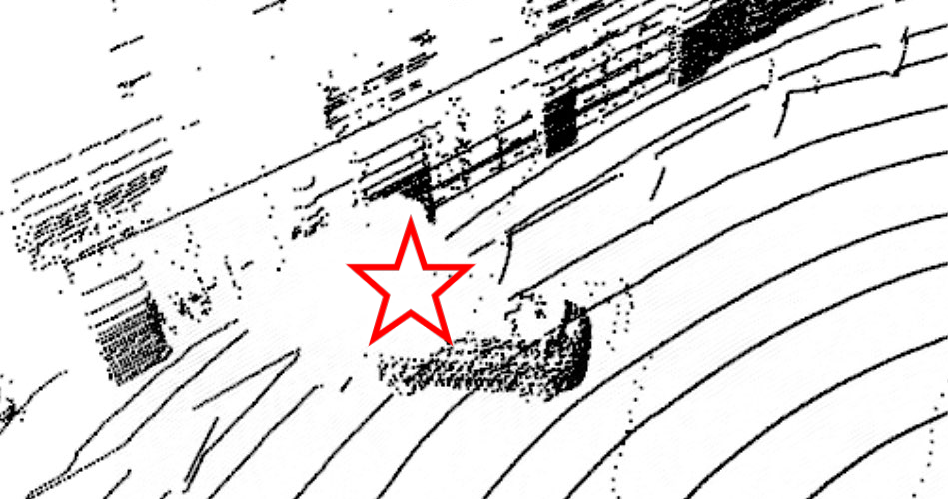}
    }
    \\ 
    \subfigure[No Collaboration]{
    	\includegraphics[width=1.6in]{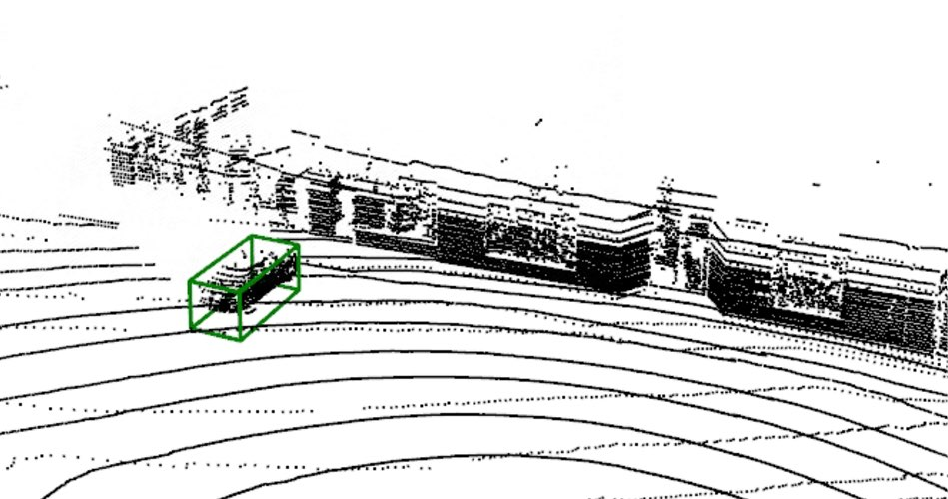}
    }
    \subfigure[Collaboration]{
	\includegraphics[width=1.6in]{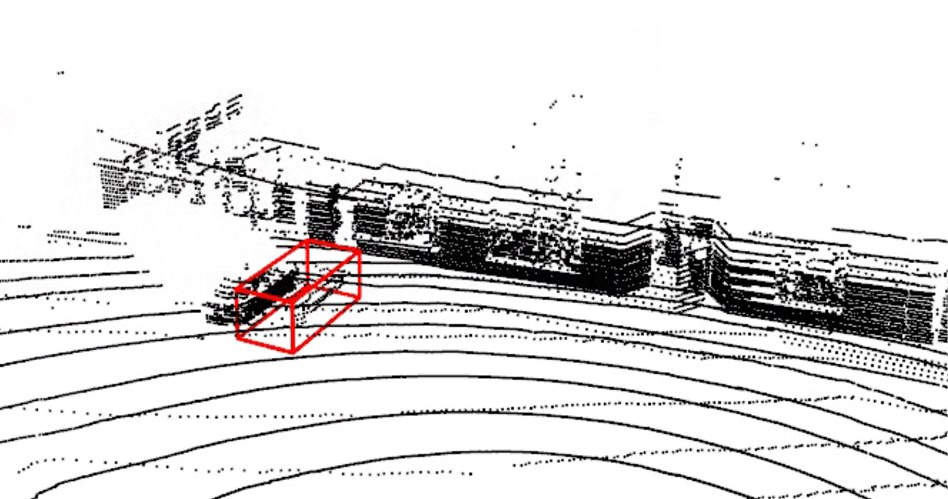}
    }
    \subfigure[No Collaboration]{
	\includegraphics[width=1.6in]{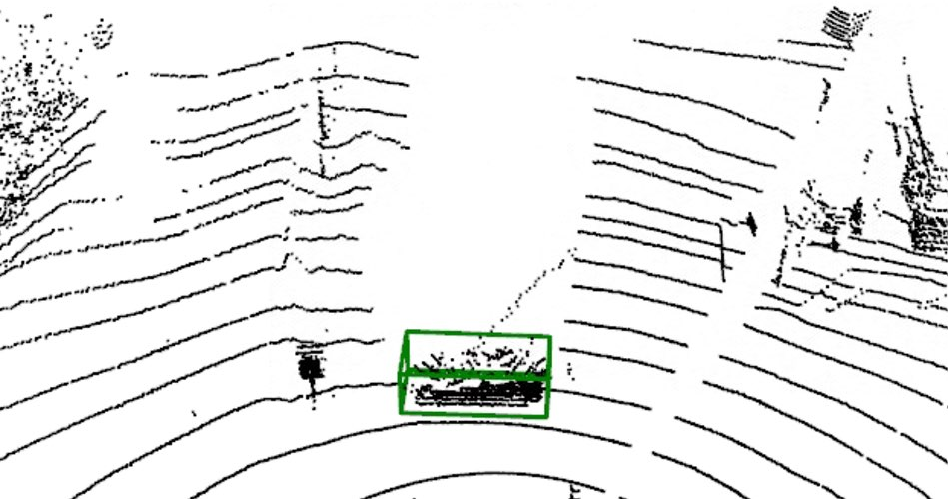}
    }
    \subfigure[Collaboration]{
	\includegraphics[width=1.6in]{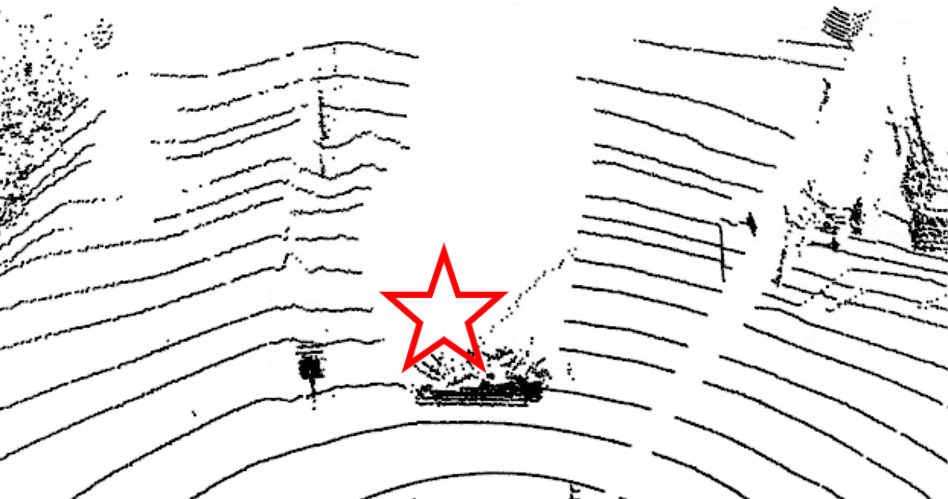}
    }
    \\
    \subfigure[No Collaboration]{
        \includegraphics[width=1.6in]{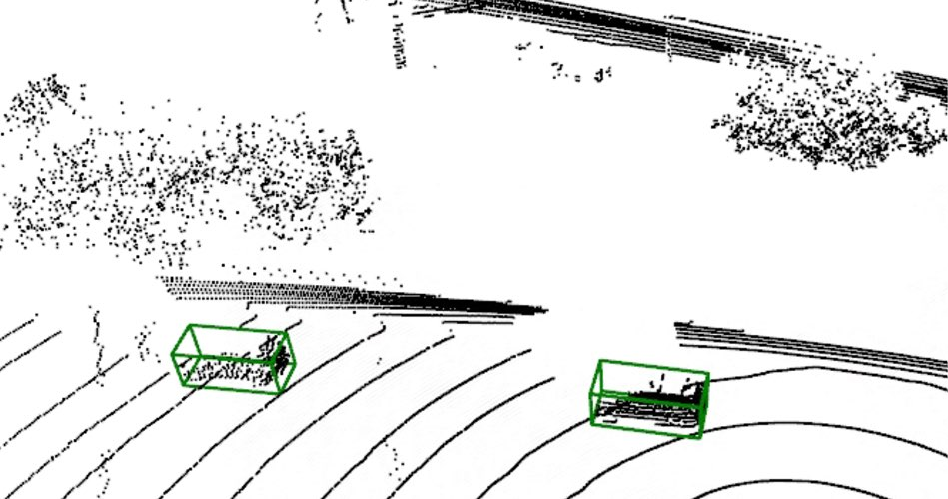}
    }
    \subfigure[Collaboration]{
	\includegraphics[width=1.6in]{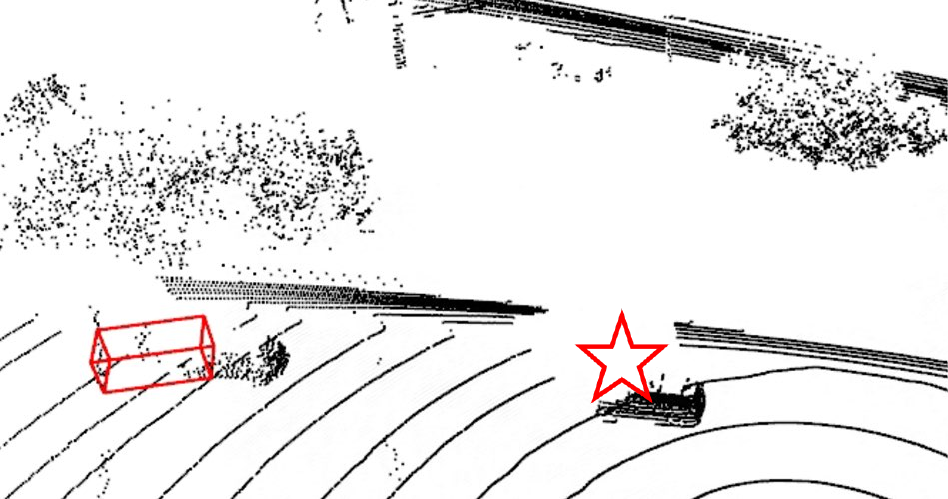}
    }
    \subfigure[No Collaboration]{
	\includegraphics[width=1.6in]{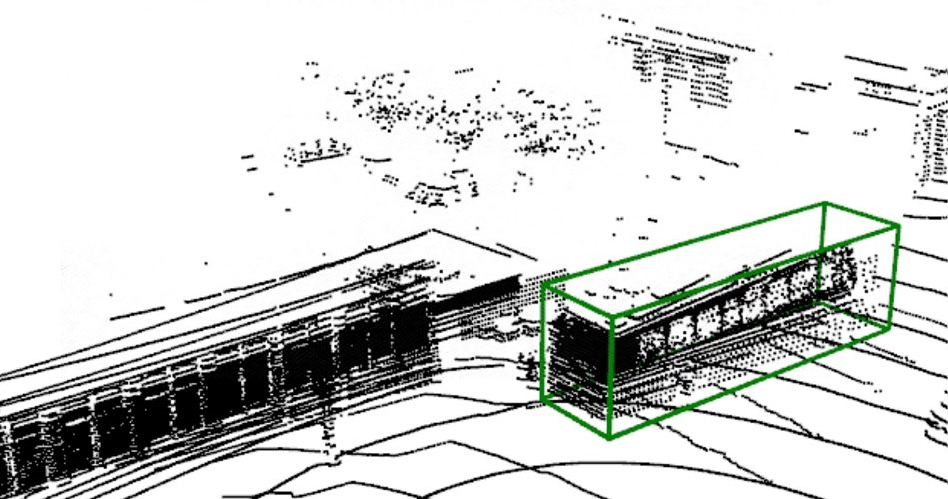}
    }
    \subfigure[Collaboration]{
	\includegraphics[width=1.6in]{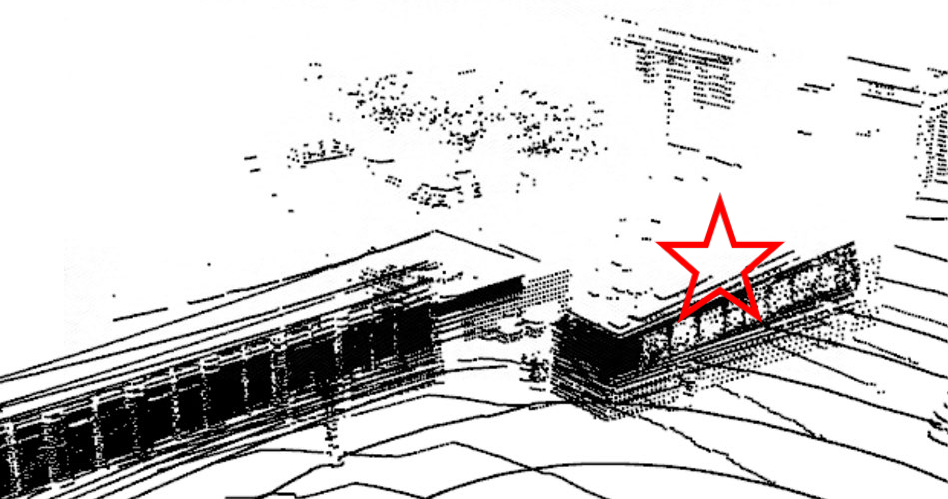}
    }
    \vspace{-5pt}
    \caption{Visualization samples showing misleading cooperation errors detected by CooTest using transformed scenes. The cars detected by the ego vehicle are marked in green, while the cars detected by cooperative perception systems are marked in red. }
    \label{fig.visual}
    \vspace{-5pt}
\end{figure*}

Then we employ the AP, which is widely utilized in object detection research for autonomous driving~\cite{padilla2020survey}, to measure the testing performance of CooTest.
To achieve fair comparison results, we use the 11 recall positions proposed in the Pascal VOC benchmark~\cite{everingham2010pascal}, which is defined as follows:

\begin{equation}
\left.\mathrm{AP}\right|_{R}=\frac{1}{|R|} \sum_{r \in R} \rho_{\text {interp }}(r)
\end{equation}

We apply exactly 11 equally spaced recall levels, i.e., $R_{11}=\{0, 0.1, 0.2, \ldots, 1\}$. The interpolation function is defined as $\rho_{\text {interp }}(r)=\max _{r^{\prime}: r^{\prime} \geq r} \rho\left(r^{\prime}\right)$, where $\rho(r)$ gives the precision at recall $r$. A higher AP means a better performance of cooperative perception systems.

\begin{table}[htbp]\footnotesize
  \centering
  \caption{The AP of different models on the transformed testing sets under different perceiving ranges.}
  \setlength{\tabcolsep}{2.5pt}
  \renewcommand\arraystretch{1.1}
    \begin{tabular}{c|cc|cc|cc|cc}
    \hline
    \multirow{2}[4]{*}{\textbf{Model}} & \multicolumn{2}{c|}{\textbf{Short}} & \multicolumn{2}{c|}{\textbf{Middle}} & \multicolumn{2}{c|}{\textbf{Long}} & \multicolumn{2}{c}{\textbf{Overall}} \bigstrut\\
\cline{2-9}          & \textbf{Before} & \textbf{After} & \textbf{Before} & \textbf{After} & \textbf{Before} & \textbf{After} & \textbf{Before} & \textbf{After} \bigstrut\\
    \hline
    \textbf{Early Fusion} & 76.1  & \textbf{66.1} & 42.4  & \textbf{29.6} & 47.7  & \textbf{36.7} & 59.7  & \textbf{49.6} \bigstrut\\
    \hline
    \textbf{Late Fusion} & 73.5  & \textbf{68.2} & 43.7  & \textbf{34.5} & 36.3  & \textbf{31.1} & 55.1  & \textbf{48.7} \bigstrut\\
    \hline
    \textbf{V2VNet} & 80.6  & \textbf{72.1} & 52.6  & \textbf{34.9} & 42.6  & \textbf{26.8} & 64.6  & \textbf{50.9} \bigstrut\\
    \hline
    \textbf{V2X-ViT} & 82    & \textbf{72.7} & 51.7  & \textbf{36} & 43.2  & \textbf{28.9} & 65    & \textbf{53.5} \bigstrut\\
    \hline
    \textbf{AttFusion} & 79.8  & \textbf{72.3} & 53.1  & \textbf{38.2} & 43.6  & \textbf{32} & 64.7  & \textbf{54.5} \bigstrut\\
    \hline
    \textbf{F-Cooper} & 80.8  & \textbf{60.4} & 45.6  & \textbf{23.3} & 32.8  & \textbf{14.9} & 60.7  & \textbf{45.4} \bigstrut\\
    \hline
    \end{tabular}%
  \label{tab:rq1}%
  \vspace{-10pt}
\end{table}%

\section{Result Analysis and Discussion}
\subsection{Answer to RQ1}

To verify whether the transformed test dataset can effectively detect the erroneous behaviors, CooTest applies seven transformations introduced in Section \ref{section:transformation operators} for each cooperative perception model on the original test set $T_{O}$ to generate test data. To avoid the inconsistency between the size of the transformed test dataset $T_{A}$ and the original test dataset $T_{O}$, CooTest generates the same size of the transformed dataset as $T_{O}$ for each transformation operator. Then the transformed datasets $T_{A}=\{T_{CT}, T_{SM}, T_{GL}, T_{CL}, T_{RN}, T_{SW}, T_{FG}\}$ are put into the pre-trained model $M$ to obtain 3D bounding box prediction outputs. Then we calculate the AP of car detection and total misleading cooperation errors according to the prediction results. We further analyze the impact of these transformation operators on the performance of cooperative perception systems under different perceiving ranges. 
Specifically, we divided the detection target into three ranges~(refer to~\cite{xu2023v2v4real,liu2023towards}) based on the distance relative to the ego vehicle, including 0~\textasciitilde 30m~(Short), 30~\textasciitilde 50m~(Middle), and 50~\textasciitilde 100m~(Long). Then we calculate the AP value on all cooperative models for each level respectively.
The drop of AP on transformed test sets indicates that CooTest can effectively detect the potential defects of the model under test.


\textbf{Results.}
Figure~\ref{fig:rq1} shows the AP of all tested models on the seed sets $T_{O}$ and transformed testing sets $T_{A}$. The AP of all tested models decreases significantly on most transformed testing sets. The AP decline percentages of the models after applying transformation operators are 20.6\% (RN), 21.7\% (SW), 22.3\% (FG), 16\% (CL), 20.3\% (GL), 8.3\% (CT), and 1.8\% (SM), respectively. Therefore, both lossy communication and adverse weather can greatly impair the accuracy of collaborative perception models. Table~\ref{tab:rq1} further elaborates the average precision under different perceiving ranges before and after applying operators. After applying transformation operators, the AP decreases for short, middle, and long-distance cooperative perception are 10.2\%, 15.4\%, and 12.6\%, respectively. These results show that all the proposed transformation operators have a positive effect on the model testing process.



\begin{table}[htbp]\footnotesize
  \centering
  \caption{The AP and MCE of models on the transformed testing sets generated with different guidance approaches.}
    \begin{tabular}{c|cc|cc|cc|cc}
    \hline
    \multirow{2}[4]{*}{\textbf{Model}} & \multicolumn{2}{c|}{\textbf{VGT(10\%)}} & \multicolumn{2}{c|}{\textbf{Ran.(10\%)}} & \multicolumn{2}{c|}{\textbf{VGT(15\%)}} & \multicolumn{2}{c}{\textbf{Ran.(15\%)}} \bigstrut\\
\cline{2-9}          & \textbf{AP} & \textbf{MCE} & \textbf{AP} & \textbf{MCE} & \textbf{AP} & \textbf{MCE} & \textbf{AP} & \textbf{MCE} \bigstrut\\
    \hline
    \textbf{Early Fusion} & \textbf{30.8} & \textbf{1030} & 51.1  & 723   & \textbf{38.7} & \textbf{1670} & 51.5  & 989 \bigstrut\\
    \hline
    \textbf{Late Fusion} & \textbf{43.6} & \textbf{1202} & 48.7  & 454   & \textbf{45.0} & \textbf{1488} & 48.4  & 652 \bigstrut\\
    \hline
    \textbf{V2VNet} & \textbf{15.9} & \textbf{587} & 53.1  & 443   & \textbf{25.4} & \textbf{1100} & 51.6  & 668 \bigstrut\\
    \hline
    \textbf{V2X-ViT} & \textbf{43.8} & \textbf{756} & 54.6  & 416   & \textbf{46.3} & \textbf{1300} & 54.5  & 626 \bigstrut\\
    \hline
    \textbf{AttFusion} & \textbf{43.3} & \textbf{935} & 54.9  & 412   & \textbf{46.1} & \textbf{1416} & 55.5  & 586 \bigstrut\\
    \hline
    \textbf{F-Cooper} & \textbf{39.5} & \textbf{1844} & 46.0    & 1210  & \textbf{41.6} & \textbf{2433} & 46.5  & 1841 \bigstrut\\
    \hline
    \end{tabular}%
  \label{tab:rq2}%
  \vspace{-5pt}
\end{table}%

\textbf{Discussion.} 
According to the results of the analysis, transformation operators can reduce the AP of all models within a certain range.
Specifically, operators have a minor impact on the perception of close objects and a more significant impact on the perception of distant objects due to the sparse nature of the transformed point cloud.
Moreover, the experimental results show that the same transformation operator has different effects on various cooperative perception models.
Based on the observations in Figure ~\ref{fig:rq1}, the late fusion strategy demonstrates greater resistance to interference from weather conditions and communication noise compared to other fusion schemes. 
We also find \textit{V2X-ViT} and \textit{AttFusion} demonstrate superior robustness to lossy communication operators (including CL and GL) compared to the other two intermediate fusion models. This resilience may be attributed to the presence of attention modules that enable them to handle shared lossy information effectively.  




\subsection{Answer to RQ2}
To evaluate the effectiveness of the proposed V2X-oriented guided transformation~(VGT) in detecting erroneous behaviors, we select the random strategy as the baseline.
Specifically, we divide the original test set into two halves ($T_{h1}$ and $T_{h2}$), using one half ($T_{h1}$) as the initial seeds to generate transformed test cases through both random testing and V2X-oriented guided testing. In the experiments, we apply the random method and the V2X-guided transformation method to select and retain the same batch of generated data. For each guided strategy, we retain 10\% and 15\% of the transformed data to evaluate the effectiveness of the guidance strategies.
Then we calculate the AP in generated test cases of cooperative perception models under test for each configuration. Furthermore, we count the number of misleading cooperation errors (see Section~\ref{section:definition}) and visualize some typical instances of misleading cooperation errors detected by CooTest to provide a direct understanding.

\begin{table*}[htbp]\small
  \centering
  \caption{The test result comparison after retraining the model with the transformed data generated by CooTest.}
  \setlength{\tabcolsep}{3pt}
  \renewcommand\arraystretch{1}
    \begin{tabular}{c|cc|cc|cc|cc|cc|cc|cc|cc}
    \hline
    \multirow{3}[6]{*}{\textbf{Model}} & \multicolumn{4}{c|}{\textbf{Short}} & \multicolumn{4}{c|}{\textbf{Middle}} & \multicolumn{4}{c|}{\textbf{Long}} & \multicolumn{4}{c}{\textbf{Overall}} \bigstrut\\
\cline{2-17}          & \multicolumn{2}{c|}{\textbf{AP}} & \multicolumn{2}{c|}{\textbf{MCE}} & \multicolumn{2}{c|}{\textbf{AP}} & \multicolumn{2}{c|}{\textbf{MCE}} & \multicolumn{2}{c|}{\textbf{AP}} & \multicolumn{2}{c|}{\textbf{MCE}} & \multicolumn{2}{c|}{\textbf{AP}} & \multicolumn{2}{c}{\textbf{MCE}} \bigstrut\\
\cline{2-17}          & \textbf{Before} & \textbf{After} & \textbf{Before} & \textbf{After} & \textbf{Before} & \textbf{After} & \textbf{Before} & \textbf{After} & \textbf{Before} & \textbf{After} & \textbf{Before} & \textbf{After} & \textbf{Before} & \textbf{After} & \textbf{Before} & \textbf{After} \bigstrut\\
    \hline
    \textbf{Early Fusion} & 65.9  & \textbf{71.4 } & 3708  & \textbf{1974 } & 30.0  & \textbf{35.6 } & 2716  & \textbf{1638 } & 36.5  & \textbf{38.7 } & 613   & \textbf{354 } & 49.5  & \textbf{53.3 } & 7037  & \textbf{3966 } \bigstrut\\
    \hline
    \textbf{Late Fusion} & 68.6  & \textbf{73.5 } & 2326  & \textbf{2001 } & 35.0  & \textbf{47.0 } & 1718  & \textbf{1078 } & 31.5  & \textbf{47.4 } & 538   & \textbf{317 } & 49.4  & \textbf{60.6 } & 4582  & \textbf{3396 } \bigstrut\\
    \hline
    \textbf{V2VNet} & 71.5  & \textbf{79.0 } & 1810  & \textbf{655 } & 34.4  & \textbf{42.3 } & 2070  & \textbf{913 } & 27.0  & \textbf{31.4 } & 613   & \textbf{373 } & 50.7  & \textbf{57.0 } & 4493  & \textbf{1941 } \bigstrut\\
    \hline
    \textbf{V2X-ViT} & 71.9  & \textbf{79.8 } & 1434  & \textbf{513 } & 35.9  & \textbf{45.4 } & 1985  & \textbf{719 } & 29.3  & \textbf{33.3 } & 696   & \textbf{366 } & 53.3  & \textbf{59.2 } & 4115  & \textbf{1598 } \bigstrut\\
    \hline
    \textbf{AttFusion} & 71.4  & \textbf{78.8} & 1874  & \textbf{1131 } & 37.8  & \textbf{45.0} & 1648  & \textbf{1111 } & 32.3  & \textbf{35.6} & 632   & \textbf{458 } & 53.9  & \textbf{60.2} & 4154  & \textbf{2700 } \bigstrut\\
    \hline
    \textbf{F-Cooper} & 59.9  & \textbf{71.3} & 7425  & \textbf{1313} & 23.7  & \textbf{37.5} & 3784  & \textbf{1217} & 15.4  & \textbf{29.9} & 1084  & \textbf{424} & 45.4  & \textbf{52.0} & 12293 & \textbf{2954} \bigstrut\\
    \hline
    \end{tabular}%
  \label{tab:rq3}%
\end{table*}%

\textbf{Results.} Table~\ref{tab:rq2} presents the testing results of different models on the tests generated by two guidance approaches. Obviously, compared to the random guidance strategy, the proposed V2X-oriented guidance method can help to find more misleading cooperation errors in all configurations.
Specifically, V2X-oriented guidance identifies at least 32.2\% more misleading cooperation errors compared to random guidance for each configuration. 
Furthermore, we find that a large number of misleading cooperation errors could lead to a sharp decline in AP. On average, V2X-oriented guidance proves to be 29.2\% and 20.9\% more effective than random guidance when retaining 10\% and 15\% of transformed data, respectively.
These results underscore that CooTest can achieve higher testing efficiency by utilizing the V2X-oriented guided transformation method.


To prove the effectiveness of CooTest in detecting misleading cooperation errors, we intercept some motivating examples found by CooTest, as shown in Figure~\ref{fig.visual}. We provide a list of six real examples of misleading cooperation errors in the model under test, encompassing diverse and challenging communication scenes.
As depicted in Figure~\ref{fig.visual}, CooTest is capable of identifying misleading cooperation errors covering multiple object types (e.g., car, bus, etc.) in various transformation scenes.
Specifically, surrounding objects accurately perceived by the ego vehicle are marked with green cuboids.
However, when involving connected vehicle collaboration, incorrect perception predictions are made, including inaccurate position estimation (marked with the red cuboid) and the inability to detect ground truth objects (marked with the red five-pointed star).
One possible reason lies in that the ego vehicle's perception system prioritizes inaccurate shared information from collaborators, even when the ego vehicle has correct environmental understanding. 


\textbf{Discussion.} To investigate the effectiveness of misleading cooperation errors found by CooTest, we conduct an analysis of the number of misleading cooperation errors and AP values identified by the V2X-guided transformation method. The analysis results reveal that misleading cooperation errors can adversely affect the performance of the cooperative perception model. Additionally, our emphasis on misleading cooperation errors stems from the intuitive expectation that the cooperative perception system should aid the ego vehicle in detecting occlusion and distant objects even in various adverse communication scenarios. However, existing cooperative perception systems struggle to distinguish misleading information even if the ego vehicle accurately perceives its surroundings. For example, from Figure~\ref{fig.visual}(g) and Figure~\ref{fig.visual}(h), it is evident that the ego vehicle correctly detects objects in challenging snowy scenes. However, it fails to detect the ground truth object after collaborating with connected autonomous vehicles. We posit that mitigating such errors will lead to a further enhancement in the perception performance of the cooperative perception system. 





\subsection{Answer to RQ3}
Here we investigate whether retraining the cooperative perception models with test scenes generated by CooTest helps enhance the model's robustness. 
We apply the generated test suites in RQ2 for model training.
To demonstrate the improvement of robustness clearly, we leverage the remaining data $T_{h2}$ (introduced in RQ2) to construct the validation set.
Specifically, we leverage each transform operator to generate a transformed data set with the same size as $T_{h2}$, and then combine all these transformed data to create a large-scale validation set.
We keep the experimental configuration of retraining consistent with the initial training model process, and the retraining epoch is set as 5. To further ensure that the retrained models perform effectively in head scenes, we tested all retrained models on the original test scenes.


\textbf{Results.} Table~\ref{tab:rq3} presents the AP and misleading cooperation error numbers under different perceiving ranges of all models after retraining with the transformed data generated by CooTest. As shown in Table~\ref{tab:rq3}, it is evident that regardless of the fusion scheme employed by the cooperative perception model, CooTest can consistently enhance the AP and mitigate misleading cooperation errors across varying perceiving ranges after retraining. Compared with the perception performance before retraining, the overall improvement range of AP is 7.7\% \textasciitilde 22.7\%, and the reduction percentage range of misleading cooperation error is 25.9\% \textasciitilde 76\%. Additionally, experiments show that among the six models, only the F-cooper model exhibits a slight decrease in performance in head scenes. In contrast, the performance of the other models remains almost the same or slightly improved. Overall, the average accuracy of all retrained models in the original test scenes increased by 3.1\%.

\textbf{Discussion.} 
Our experiments show that using transformed data generated by CooTest for retraining can significantly improve the model robustness on the cooperative perception task. 
Meanwhile, we find it possible to reduce misleading cooperation errors by leveraging retraining techniques to increase the cooperative perception system's overall performance. 
However, we find that the retrained model still suffers from a variable number of misleading cooperation errors.
One possible reason is that cooperative perception systems involve multiple agents collaborating in complex traffic scenarios.
The complex model structure and specific communication mechanisms make it difficult to directly apply retraining techniques to completely repair all cooperative perception errors.
Considering the significant impact of misleading cooperation errors on the system, more research on the continuous enhancement of the cooperative perception system, such as debugging and repairing, is needed.

\subsection{Threats to Validity}

\textbf{\emph{Data Selection.}} Data selection stands as one of the fundamental threats to validity. The limitation of the experiment dataset may threaten the generality of the results. The quality of the transformed data is fundamentally contingent on the characteristics of the original dataset. Consequently, when it comes to input data featuring driving scenes not present in the training set, the retrained model may face challenges in accurate predictions. To alleviate this threat, CooTest adopts a diverse range of transformation operators and a widely used large-scale dataset for experiments.

\noindent\textbf{\emph{Data Simulation.}} 
One threat to validity comes from the data simulation. On the one hand, various additional factors can potentially introduce communication issues, thereby influencing the behaviors of cooperative perception systems. On the other hand, the generated data may not be perfectly reproducible in the real world due to a multitude of unpredictable factors (e.g., the back-scattering coefficient of the target, the parameters of the LiDAR sensors, etc). However, our transformations are meticulously designed to be realistic, introduced in Section~\ref{section:4.2}. Besides, we take into consideration unexpected environmental conditions and changes for V2X communication systems to the greatest extent possible.


\noindent\textbf{\emph{Parameter Settings.}} An additional potential threat may stem from the variability in configurable transformation parameters. Some of our findings might not necessarily extend to different parameter sets during the evaluation phase. To mitigate this concern, we meticulously selected a wide array of parameters that closely replicate the actual cooperative driving environment within our experimental design. We delve into the candidate parameters for each transformation in Section~\ref{section:4.2}.

\noindent\textbf{\emph{The Testing Functionality Selection.}}  
CooTest conducts only offline testing~\cite{DBLP:journals/ese/StoccoPT23,haq2021can} and identifies cooperative issues in LiDAR-based cooperative object detection. While these issues are significant for perception module developers, this study does not assess their impact on subsequent tasks such as decision-making, planning, and control. This limitation underscores a crucial direction for future research.


\section{Related Work}


\subsection{Testing and Verification of Autonomous Driving Perception Systems}

As a foundational component in supporting ADS industrial applications, the ability of perception systems to understand the environment significantly influences the overall system quality~\cite{gruyer2017perception,li2020lidar}. Recent research efforts have primarily concentrated on robustness testing of single-agent perception systems across various driving tasks~\cite{wang2020metamorphic,xie2022towards,guo2022lirtest,christian2023generating,gao2023benchmarking}. Wang et al.~\cite{wang2020metamorphic} propose MetaOD to generate data
by inserting object instances into the background image to test camera-based perception systems. Then they assert equality criteria between the original and synthesized images to reveal object detection model defects. Moreover, Xie et al.~\cite{xie2022towards} effectively assess the robustness of object tracking systems in scenarios characterized by varying speeds of tracked objects and temporary camera failures. 
There are also a few works that focus on the testing of LiDAR-based perception systems.
LiRTest~\cite{guo2022lirtest} utilizes metamorphic relations to generate testing point clouds and identifies failures in the 3D object detection models. Christian et al.~\cite{christian2023generating} propose an innovative approach for testing LiDAR-based perception systems by applying semantic mutations to labeled real-world data.

The key difference between the aforementioned related works and CooTest lies in their focus on single-vehicle perception quality assurance techniques, whereas our approach involves designing various transformation operators specifically for testing the V2X-oriented cooperative perception module. We are the first to establish the concept of faults attributed to misleading cooperation errors and formulate a V2X-oriented guided transformation strategy to improve the test efficiency of cooperative perception systems.


\subsection{Metamorphic Testing}

Metamorphic Testing (MT)~\cite{segura2016survey,chen2018metamorphic} is a software testing methodology that alleviates the test oracle problem. It discerns software bugs by detecting violations of domain-specific MRs defined across outputs from multiple executions of the program with different test inputs. In conventional software testing, researchers propose the utilization of MT to debugging~\cite{jin2015concolic}, proving~\cite{2002-Semi-proving}, fault localization~\cite{xie2011spectrum}, program slicing~\cite{xie2013metamorphic}, program repair~\cite{jiang2017metamorphic}, etc. 



With the swift advancement of deep learning, metamorphic testing has found application in testing both classification and regression tasks~\cite{guo2024semantic, xie2011testing, wang2020metamorphic, DBLP:journals/tosem/ZhangZFLSHW24}. Xie et al.~\cite{xie2011testing} introduce an MT technique to test machine learning classification algorithms, demonstrating its effectiveness in identifying faults within a widely recognized open-source classification program. Besides, Wang et al.~\cite{wang2020metamorphic} employs MT as an adaptive and effective testing strategy by asserting the equivalence of object detection results between original and synthetic images to expose defects in object detection systems. Guo et al.~\cite{guo2022lirtest} introduce LiDAR-based specific metamorphic relations and employ them to generate various corner test cases.


The metamorphic testing methods mentioned above cannot be directly applied in the V2X field due to the unique communication characteristics of cooperative perception systems. In this paper, CooTest develops a specific V2X-oriented metamorphic testing module for cooperative perception systems, where the communication and weather MRs are defined in Section~\ref{section:MR}. As a result, erroneous cooperative perception results can be unveiled by checking the failure of these MRs.

\section{Conclusion}

In this paper, we propose and evaluate CooTest, the first systematic and automated module testing tool for V2X communication systems. In order to enrich the test samples and uncover corner cases for the models under test, CooTest has introduced seven innovative transformation operators. These operators are utilized to generate synthetic test data by applying transformations to seed data. We use V2X domain-specific MRs to find misleading cooperation errors in the cooperative perception models without detailed specifications.

CooTest has been evaluated using six state-of-the-art cooperative detection models employing various fusion methods. The experimental outcomes demonstrated that the transformed test data generated by CooTest efficiently detected erroneous behaviors and notably enhanced the robustness of V2X cooperative perception models through retraining. This substantiates CooTest's significant role in ensuring the quality and reliability of autonomous driving perception systems.

\section{Data Availability}

Our data are available on the website: \url{https://github.com/meng2180/CooTest}


\bibliographystyle{ACM-Reference-Format}
\balance
\bibliography{reference}


\begin{thebibliography}{60}


\ifx \showCODEN    \undefined \def \showCODEN     #1{\unskip}     \fi
\ifx \showDOI      \undefined \def \showDOI       #1{#1}\fi
\ifx \showISBNx    \undefined \def \showISBNx     #1{\unskip}     \fi
\ifx \showISBNxiii \undefined \def \showISBNxiii  #1{\unskip}     \fi
\ifx \showISSN     \undefined \def \showISSN      #1{\unskip}     \fi
\ifx \showLCCN     \undefined \def \showLCCN      #1{\unskip}     \fi
\ifx \shownote     \undefined \def \shownote      #1{#1}          \fi
\ifx \showarticletitle \undefined \def \showarticletitle #1{#1}   \fi
\ifx \showURL      \undefined \def \showURL       {\relax}        \fi
\providecommand\bibfield[2]{#2}
\providecommand\bibinfo[2]{#2}
\providecommand\natexlab[1]{#1}
\providecommand\showeprint[2][]{arXiv:#2}

\bibitem[aut(2015)]%
        {autoware}
 \bibinfo{year}{2015}\natexlab{}.
\newblock \bibinfo{title}{Autoware}.
\newblock
\newblock
\urldef\tempurl%
\url{https://github.com/Autoware-AI/autoware.ai}
\showURL{%
\tempurl}


\bibitem[apo(2017)]%
        {apollo}
 \bibinfo{year}{2017}\natexlab{}.
\newblock \bibinfo{title}{Apollo}.
\newblock
\newblock
\urldef\tempurl%
\url{https://github.com/ApolloAuto/apollo}
\showURL{%
\tempurl}


\bibitem[cra(2018)]%
        {crash-bicycle1}
 \bibinfo{year}{2018}\natexlab{}.
\newblock \bibinfo{title}{LIDAR Maker Velodyne Shifts Away Blame In Fatal Uber
  Self-Driving Crash}.
\newblock
  \bibinfo{howpublished}{\url{https://jalopnik.com/lidar-maker-velodyne-blame-to-uber-in-fatal-self-drivin-1824027977}}.
\newblock


\bibitem[for(2019)]%
        {ford-v2x}
 \bibinfo{year}{2019}\natexlab{}.
\newblock \bibinfo{title}{FORD ACCELERATES CONNECTIVITY STRATEGY IN CHINA;
  TARGETS PRODUCTION OF FIRST C-V2X-EQUIPPED VEHICLE IN 2021}.
\newblock
  \bibinfo{howpublished}{\url{https://media.ford.com/content/fordmedia/fap/cn/en/news/2019/03/26/Ford_Accelerates_Connectivity_Strategy_in_China_And_Targets_Production_of_First_C-V2X-Equipped_Vehicle_in_2021.html}}.
\newblock


\bibitem[cra(2019)]%
        {crash}
 \bibinfo{year}{2019}\natexlab{}.
\newblock \bibinfo{title}{Tesla’s Autopilot was engaged when Model 3 crashed
  into truck, report states}.
\newblock
\newblock
\urldef\tempurl%
\url{https://www.theverge.com/2019/5/16/18627766/tesla-autopilot-fatal-crash-delray-florida-ntsb-model-3}
\showURL{%
\tempurl}


\bibitem[apo(2023)]%
        {apollo-v2x}
 \bibinfo{year}{2023}\natexlab{}.
\newblock \bibinfo{title}{ASTRI and Baidu Apollo Collaborate to Promote the
  Implementation of C-V2X Technology in Hong Kong}.
\newblock
  \bibinfo{howpublished}{\url{https://www.astri.org/news-detail/astri-and-baidu-apollo-collaborate-to-promote-the-implementation-of-c-v2x-technology-in-hong-kong/}}.
\newblock


\bibitem[Belyaev et~al\mbox{.}(2015)]%
        {belyaev2014robust}
\bibfield{author}{\bibinfo{person}{Evgeny Belyaev}, \bibinfo{person}{Alexey~V.
  Vinel}, \bibinfo{person}{Adam Surak}, \bibinfo{person}{Moncef Gabbouj},
  \bibinfo{person}{Magnus Jonsson}, {and} \bibinfo{person}{Karen~O.
  Egiazarian}.} \bibinfo{year}{2015}\natexlab{}.
\newblock \showarticletitle{Robust Vehicle-to-Infrastructure Video Transmission
  for Road Surveillance Applications}.
\newblock \bibinfo{journal}{\emph{{IEEE} Trans. Veh. Technol.}}
  \bibinfo{volume}{64}, \bibinfo{number}{7} (\bibinfo{year}{2015}),
  \bibinfo{pages}{2991--3003}.
\newblock
\urldef\tempurl%
\url{https://doi.org/10.1109/TVT.2014.2354376}
\showDOI{\tempurl}


\bibitem[Chen et~al\mbox{.}(2022)]%
        {chen2022milestones}
\bibfield{author}{\bibinfo{person}{Long Chen}, \bibinfo{person}{Yuchen Li},
  \bibinfo{person}{Chao Huang}, \bibinfo{person}{Bai Li}, \bibinfo{person}{Yang
  Xing}, \bibinfo{person}{Daxin Tian}, \bibinfo{person}{Li Li},
  \bibinfo{person}{Zhongxu Hu}, \bibinfo{person}{Xiaoxiang Na},
  \bibinfo{person}{Zixuan Li}, {et~al\mbox{.}}}
  \bibinfo{year}{2022}\natexlab{}.
\newblock \showarticletitle{Milestones in autonomous driving and intelligent
  vehicles: Survey of surveys}.
\newblock \bibinfo{journal}{\emph{IEEE Transactions on Intelligent Vehicles}}
  \bibinfo{volume}{8}, \bibinfo{number}{2} (\bibinfo{year}{2022}),
  \bibinfo{pages}{1046--1056}.
\newblock
\urldef\tempurl%
\url{https://doi.org/10.1109/TIV.2022.3223131}
\showDOI{\tempurl}


\bibitem[Chen et~al\mbox{.}(2019a)]%
        {2019-F-cooper}
\bibfield{author}{\bibinfo{person}{Qi Chen}, \bibinfo{person}{Xu Ma},
  \bibinfo{person}{Sihai Tang}, \bibinfo{person}{Jingda Guo},
  \bibinfo{person}{Qing Yang}, {and} \bibinfo{person}{Song Fu}.}
  \bibinfo{year}{2019}\natexlab{a}.
\newblock \showarticletitle{F-cooper: feature based cooperative perception for
  autonomous vehicle edge computing system using 3D point clouds}. In
  \bibinfo{booktitle}{\emph{Proceedings of the 4th Symposium on Edge
  Computing}}. \bibinfo{publisher}{{ACM}}, \bibinfo{address}{Arlington,
  Virginia, USA}, \bibinfo{pages}{88--100}.
\newblock
\urldef\tempurl%
\url{https://doi.org/10.1145/3318216.3363300}
\showDOI{\tempurl}


\bibitem[Chen et~al\mbox{.}(2019b)]%
        {2019-Cooper}
\bibfield{author}{\bibinfo{person}{Qi Chen}, \bibinfo{person}{Sihai Tang},
  \bibinfo{person}{Qing Yang}, {and} \bibinfo{person}{Song Fu}.}
  \bibinfo{year}{2019}\natexlab{b}.
\newblock \showarticletitle{Cooper: Cooperative Perception for Connected
  Autonomous Vehicles Based on 3D Point Clouds}. In
  \bibinfo{booktitle}{\emph{Proceedings of the 39th International Conference on
  Distributed Computing Systems}}. \bibinfo{publisher}{{IEEE}},
  \bibinfo{address}{Dallas, TX, USA}, \bibinfo{pages}{514--524}.
\newblock
\urldef\tempurl%
\url{https://doi.org/10.1109/ICDCS.2019.00058}
\showDOI{\tempurl}


\bibitem[Chen et~al\mbox{.}(2020)]%
        {2020-Metamorphic-Testing}
\bibfield{author}{\bibinfo{person}{Tsong~Yueh Chen}, \bibinfo{person}{S.~C.
  Cheung}, {and} \bibinfo{person}{Siu{-}Ming Yiu}.}
  \bibinfo{year}{2020}\natexlab{}.
\newblock \showarticletitle{Metamorphic Testing: {A} New Approach for
  Generating Next Test Cases}.
\newblock \bibinfo{journal}{\emph{CoRR}} \bibinfo{volume}{abs/2002.12543},
  \bibinfo{number}{1} (\bibinfo{year}{2020}), \bibinfo{pages}{1--11}.
\newblock
\urldef\tempurl%
\url{https://arxiv.org/abs/2002.12543}
\showURL{%
\tempurl}


\bibitem[Chen et~al\mbox{.}(2018)]%
        {chen2018metamorphic}
\bibfield{author}{\bibinfo{person}{Tsong~Yueh Chen}, \bibinfo{person}{Fei-Ching
  Kuo}, \bibinfo{person}{Huai Liu}, \bibinfo{person}{Pak-Lok Poon},
  \bibinfo{person}{Dave Towey}, \bibinfo{person}{TH Tse}, {and}
  \bibinfo{person}{Zhi~Quan Zhou}.} \bibinfo{year}{2018}\natexlab{}.
\newblock \showarticletitle{Metamorphic testing: A review of challenges and
  opportunities}.
\newblock \bibinfo{journal}{\emph{ACM Computing Surveys (CSUR)}}
  \bibinfo{volume}{51}, \bibinfo{number}{1} (\bibinfo{year}{2018}),
  \bibinfo{pages}{1--27}.
\newblock
\urldef\tempurl%
\url{https://doi.org/10.1145/3143561}
\showDOI{\tempurl}


\bibitem[Chen et~al\mbox{.}(2002)]%
        {2002-Semi-proving}
\bibfield{author}{\bibinfo{person}{Tsong~Yueh Chen}, \bibinfo{person}{T.~H.
  Tse}, {and} \bibinfo{person}{Zhiquan Zhou}.} \bibinfo{year}{2002}\natexlab{}.
\newblock \showarticletitle{Semi-proving: an integrated method based on global
  symbolic evaluation and metamorphic testing}. In
  \bibinfo{booktitle}{\emph{Proceedings of the International Symposium on
  Software Testing and Analysis}}. \bibinfo{publisher}{{ACM}},
  \bibinfo{address}{Roma, Italy}, \bibinfo{pages}{191--195}.
\newblock
\urldef\tempurl%
\url{https://doi.org/10.1145/566172.566202}
\showDOI{\tempurl}


\bibitem[Chow et~al\mbox{.}(2021)]%
        {chow2021adaptive}
\bibfield{author}{\bibinfo{person}{Andy~HF Chow}, \bibinfo{person}{ZC Su},
  \bibinfo{person}{EM Liang}, {and} \bibinfo{person}{RX Zhong}.}
  \bibinfo{year}{2021}\natexlab{}.
\newblock \showarticletitle{Adaptive signal control for bus service reliability
  with connected vehicle technology via reinforcement learning}.
\newblock \bibinfo{journal}{\emph{Transportation Research Part C: Emerging
  Technologies}}  \bibinfo{volume}{129} (\bibinfo{year}{2021}),
  \bibinfo{pages}{103264}.
\newblock
\urldef\tempurl%
\url{https://doi.org/10.1016/j.trc.2021.103264}
\showDOI{\tempurl}


\bibitem[Christian et~al\mbox{.}(2023)]%
        {christian2023generating}
\bibfield{author}{\bibinfo{person}{Garrett Christian}, \bibinfo{person}{Trey
  Woodlief}, {and} \bibinfo{person}{Sebastian Elbaum}.}
  \bibinfo{year}{2023}\natexlab{}.
\newblock \showarticletitle{Generating Realistic and Diverse Tests for
  LiDAR-Based Perception Systems}. In \bibinfo{booktitle}{\emph{2023 IEEE/ACM
  45th International Conference on Software Engineering (ICSE)}}. IEEE,
  \bibinfo{pages}{2604--2616}.
\newblock
\urldef\tempurl%
\url{https://doi.org/10.1109/ICSE48619.2023.00217}
\showDOI{\tempurl}


\bibitem[Cui et~al\mbox{.}(2022)]%
        {cui2022cooperative}
\bibfield{author}{\bibinfo{person}{Guangzhen Cui}, \bibinfo{person}{Weili
  Zhang}, \bibinfo{person}{Yanqiu Xiao}, \bibinfo{person}{Lei Yao}, {and}
  \bibinfo{person}{Zhanpeng Fang}.} \bibinfo{year}{2022}\natexlab{}.
\newblock \showarticletitle{Cooperative perception technology of autonomous
  driving in the internet of vehicles environment: A review}.
\newblock \bibinfo{journal}{\emph{Sensors}} \bibinfo{volume}{22},
  \bibinfo{number}{15} (\bibinfo{year}{2022}), \bibinfo{pages}{5535}.
\newblock
\urldef\tempurl%
\url{https://doi.org/10.3390/S22155535}
\showDOI{\tempurl}


\bibitem[Everingham et~al\mbox{.}(2010)]%
        {everingham2010pascal}
\bibfield{author}{\bibinfo{person}{Mark Everingham}, \bibinfo{person}{Luc
  Van~Gool}, \bibinfo{person}{Christopher~KI Williams}, \bibinfo{person}{John
  Winn}, {and} \bibinfo{person}{Andrew Zisserman}.}
  \bibinfo{year}{2010}\natexlab{}.
\newblock \showarticletitle{The pascal visual object classes (voc) challenge}.
\newblock \bibinfo{journal}{\emph{International journal of computer vision}}
  \bibinfo{volume}{88} (\bibinfo{year}{2010}), \bibinfo{pages}{303--338}.
\newblock
\urldef\tempurl%
\url{https://doi.org/10.1007/S11263-009-0275-4}
\showDOI{\tempurl}


\bibitem[Fernandes et~al\mbox{.}(2021)]%
        {fernandes2021point}
\bibfield{author}{\bibinfo{person}{Duarte Fernandes},
  \bibinfo{person}{Ant{\'o}nio Silva}, \bibinfo{person}{Rafael N{\'e}voa},
  \bibinfo{person}{Cl{\'a}udia Sim{\~o}es}, \bibinfo{person}{Dibet Gonzalez},
  \bibinfo{person}{Miguel Guevara}, \bibinfo{person}{Paulo Novais},
  \bibinfo{person}{Jo{\~a}o Monteiro}, {and} \bibinfo{person}{Pedro
  Melo-Pinto}.} \bibinfo{year}{2021}\natexlab{}.
\newblock \showarticletitle{Point-cloud based 3D object detection and
  classification methods for self-driving applications: A survey and taxonomy}.
\newblock \bibinfo{journal}{\emph{Information Fusion}}  \bibinfo{volume}{68}
  (\bibinfo{year}{2021}), \bibinfo{pages}{161--191}.
\newblock
\urldef\tempurl%
\url{https://doi.org/10.1016/J.INFFUS.2020.11.002}
\showDOI{\tempurl}


\bibitem[Gao et~al\mbox{.}(2023)]%
        {gao2023benchmarking}
\bibfield{author}{\bibinfo{person}{Xinyu Gao}, \bibinfo{person}{Zhijie Wang},
  \bibinfo{person}{Yang Feng}, \bibinfo{person}{Lei Ma},
  \bibinfo{person}{Zhenyu Chen}, {and} \bibinfo{person}{Baowen Xu}.}
  \bibinfo{year}{2023}\natexlab{}.
\newblock \showarticletitle{Benchmarking Robustness of AI-enabled Multi-sensor
  Fusion Systems: Challenges and Opportunities}. In
  \bibinfo{booktitle}{\emph{Proceedings of the 31st ACM Joint European Software
  Engineering Conference and Symposium on the Foundations of Software
  Engineering}}. \bibinfo{pages}{871--882}.
\newblock
\urldef\tempurl%
\url{https://doi.org/10.1145/3611643.3616278}
\showDOI{\tempurl}


\bibitem[Gruyer et~al\mbox{.}(2017)]%
        {gruyer2017perception}
\bibfield{author}{\bibinfo{person}{Dominique Gruyer}, \bibinfo{person}{Valentin
  Magnier}, \bibinfo{person}{Karima Hamdi}, \bibinfo{person}{Laur{\`e}ne
  Claussmann}, \bibinfo{person}{Olivier Orfila}, {and} \bibinfo{person}{Andry
  Rakotonirainy}.} \bibinfo{year}{2017}\natexlab{}.
\newblock \showarticletitle{Perception, information processing and modeling:
  Critical stages for autonomous driving applications}.
\newblock \bibinfo{journal}{\emph{Annual Reviews in Control}}
  \bibinfo{volume}{44} (\bibinfo{year}{2017}), \bibinfo{pages}{323--341}.
\newblock
\urldef\tempurl%
\url{https://doi.org/10.1016/J.ARCONTROL.2017.09.012}
\showDOI{\tempurl}


\bibitem[Guo et~al\mbox{.}(2022)]%
        {guo2022lirtest}
\bibfield{author}{\bibinfo{person}{An Guo}, \bibinfo{person}{Yang Feng}, {and}
  \bibinfo{person}{Zhenyu Chen}.} \bibinfo{year}{2022}\natexlab{}.
\newblock \showarticletitle{LiRTest: augmenting LiDAR point clouds for
  automated testing of autonomous driving systems}. In
  \bibinfo{booktitle}{\emph{{ISSTA} '22: 31st {ACM} {SIGSOFT} International
  Symposium on Software Testing and Analysis, Virtual Event, South Korea, July
  18 - 22, 2022}}, \bibfield{editor}{\bibinfo{person}{Sukyoung Ryu} {and}
  \bibinfo{person}{Yannis Smaragdakis}} (Eds.). \bibinfo{publisher}{{ACM}},
  \bibinfo{pages}{480--492}.
\newblock
\urldef\tempurl%
\url{https://doi.org/10.1145/3533767.3534397}
\showDOI{\tempurl}


\bibitem[Guo et~al\mbox{.}(2024)]%
        {guo2024semantic}
\bibfield{author}{\bibinfo{person}{An Guo}, \bibinfo{person}{Yang Feng},
  \bibinfo{person}{Yizhen Cheng}, {and} \bibinfo{person}{Zhenyu Chen}.}
  \bibinfo{year}{2024}\natexlab{}.
\newblock \showarticletitle{Semantic-guided fuzzing for virtual testing of
  autonomous driving systems}.
\newblock \bibinfo{journal}{\emph{J. Syst. Softw.}}  \bibinfo{volume}{212}
  (\bibinfo{year}{2024}), \bibinfo{pages}{112017}.
\newblock
\urldef\tempurl%
\url{https://doi.org/10.1016/J.JSS.2024.112017}
\showDOI{\tempurl}


\bibitem[Han et~al\mbox{.}(2023)]%
        {han2023collaborative}
\bibfield{author}{\bibinfo{person}{Yushan Han}, \bibinfo{person}{Hui Zhang},
  \bibinfo{person}{Huifang Li}, \bibinfo{person}{Yi Jin},
  \bibinfo{person}{Congyan Lang}, {and} \bibinfo{person}{Yidong Li}.}
  \bibinfo{year}{2023}\natexlab{}.
\newblock \showarticletitle{Collaborative perception in autonomous driving:
  Methods, datasets, and challenges}.
\newblock \bibinfo{journal}{\emph{IEEE Intelligent Transportation Systems
  Magazine}} (\bibinfo{year}{2023}).
\newblock
\urldef\tempurl%
\url{https://doi.org/10.1109/MITS.2023.3298534}
\showDOI{\tempurl}


\bibitem[Haq et~al\mbox{.}(2021)]%
        {haq2021can}
\bibfield{author}{\bibinfo{person}{Fitash~Ul Haq}, \bibinfo{person}{Donghwan
  Shin}, \bibinfo{person}{Shiva Nejati}, {and} \bibinfo{person}{Lionel~Claude
  Briand}.} \bibinfo{year}{2021}\natexlab{}.
\newblock \showarticletitle{Can Offline Testing of Deep Neural Networks Replace
  Their Online Testing?}
\newblock \bibinfo{journal}{\emph{Empir. Softw. Eng.}} \bibinfo{volume}{26},
  \bibinfo{number}{5} (\bibinfo{year}{2021}), \bibinfo{pages}{90}.
\newblock
\urldef\tempurl%
\url{https://doi.org/10.1007/S10664-021-09982-4}
\showDOI{\tempurl}


\bibitem[Heinzler et~al\mbox{.}(2019)]%
        {heinzler2019weather}
\bibfield{author}{\bibinfo{person}{Robin Heinzler}, \bibinfo{person}{Philipp
  Schindler}, \bibinfo{person}{J{\"u}rgen Seekircher}, \bibinfo{person}{Werner
  Ritter}, {and} \bibinfo{person}{Wilhelm Stork}.}
  \bibinfo{year}{2019}\natexlab{}.
\newblock \showarticletitle{Weather influence and classification with
  automotive lidar sensors}. In \bibinfo{booktitle}{\emph{2019 IEEE intelligent
  vehicles symposium (IV)}}. IEEE, \bibinfo{pages}{1527--1534}.
\newblock
\urldef\tempurl%
\url{https://doi.org/10.1109/IVS.2019.8814205}
\showDOI{\tempurl}


\bibitem[Hobert et~al\mbox{.}(2015)]%
        {hobert2015enhancements}
\bibfield{author}{\bibinfo{person}{Laurens Hobert}, \bibinfo{person}{Andreas
  Festag}, \bibinfo{person}{Ignacio Llatser}, \bibinfo{person}{Luciano
  Altomare}, \bibinfo{person}{Filippo Visintainer}, {and}
  \bibinfo{person}{Andras Kovacs}.} \bibinfo{year}{2015}\natexlab{}.
\newblock \showarticletitle{Enhancements of V2X communication in support of
  cooperative autonomous driving}.
\newblock \bibinfo{journal}{\emph{IEEE communications magazine}}
  \bibinfo{volume}{53}, \bibinfo{number}{12} (\bibinfo{year}{2015}),
  \bibinfo{pages}{64--70}.
\newblock
\urldef\tempurl%
\url{https://doi.org/10.1109/MCOM.2015.7355568}
\showDOI{\tempurl}


\bibitem[Hu et~al\mbox{.}(2024)]%
        {DBLP:journals/corr/abs-2401-01544}
\bibfield{author}{\bibinfo{person}{Senkang Hu}, \bibinfo{person}{Zhengru Fang},
  \bibinfo{person}{Yiqin Deng}, \bibinfo{person}{Xianhao Chen}, {and}
  \bibinfo{person}{Yuguang Fang}.} \bibinfo{year}{2024}\natexlab{}.
\newblock \showarticletitle{Collaborative Perception for Connected and
  Autonomous Driving: Challenges, Possible Solutions and Opportunities}.
\newblock \bibinfo{journal}{\emph{CoRR}}  \bibinfo{volume}{abs/2401.01544}
  (\bibinfo{year}{2024}).
\newblock
\urldef\tempurl%
\url{https://doi.org/10.48550/ARXIV.2401.01544}
\showDOI{\tempurl}


\bibitem[Jiang et~al\mbox{.}(2017)]%
        {jiang2017metamorphic}
\bibfield{author}{\bibinfo{person}{Mingyue Jiang}, \bibinfo{person}{Tsong~Yueh
  Chen}, \bibinfo{person}{Fei{-}Ching Kuo}, \bibinfo{person}{Dave Towey}, {and}
  \bibinfo{person}{Zuohua Ding}.} \bibinfo{year}{2017}\natexlab{}.
\newblock \showarticletitle{A metamorphic testing approach for supporting
  program repair without the need for a test oracle}.
\newblock \bibinfo{journal}{\emph{J. Syst. Softw.}}  \bibinfo{volume}{126}
  (\bibinfo{year}{2017}), \bibinfo{pages}{127--140}.
\newblock
\urldef\tempurl%
\url{https://doi.org/10.1016/j.jss.2016.04.002}
\showDOI{\tempurl}


\bibitem[Jin et~al\mbox{.}(2015)]%
        {jin2015concolic}
\bibfield{author}{\bibinfo{person}{Hao Jin}, \bibinfo{person}{Yanyan Jiang},
  \bibinfo{person}{Na Liu}, \bibinfo{person}{Chang Xu},
  \bibinfo{person}{Xiaoxing Ma}, {and} \bibinfo{person}{Jian Lu}.}
  \bibinfo{year}{2015}\natexlab{}.
\newblock \showarticletitle{Concolic Metamorphic Debugging}. In
  \bibinfo{booktitle}{\emph{39th {IEEE} Annual Computer Software and
  Applications Conference, {COMPSAC} 2015, Taichung, Taiwan, July 1-5, 2015.
  Volume 2}}, \bibfield{editor}{\bibinfo{person}{Sheikh~Iqbal Ahamed},
  \bibinfo{person}{Carl~K. Chang}, \bibinfo{person}{William~C. Chu},
  \bibinfo{person}{Ivica Crnkovic}, \bibinfo{person}{Pao{-}Ann Hsiung},
  \bibinfo{person}{Gang Huang}, {and} \bibinfo{person}{Jingwei Yang}} (Eds.).
  \bibinfo{publisher}{{IEEE} Computer Society}, \bibinfo{pages}{232--241}.
\newblock
\urldef\tempurl%
\url{https://doi.org/10.1109/COMPSAC.2015.79}
\showDOI{\tempurl}


\bibitem[Jung et~al\mbox{.}(2020)]%
        {jung2020v2x}
\bibfield{author}{\bibinfo{person}{Chanyoung Jung}, \bibinfo{person}{Daegyu
  Lee}, \bibinfo{person}{Seungwook Lee}, {and} \bibinfo{person}{David~Hyunchul
  Shim}.} \bibinfo{year}{2020}\natexlab{}.
\newblock \showarticletitle{V2X-communication-aided autonomous driving: System
  design and experimental validation}.
\newblock \bibinfo{journal}{\emph{Sensors}} \bibinfo{volume}{20},
  \bibinfo{number}{10} (\bibinfo{year}{2020}), \bibinfo{pages}{2903}.
\newblock
\urldef\tempurl%
\url{https://doi.org/10.3390/S20102903}
\showDOI{\tempurl}


\bibitem[Kilic et~al\mbox{.}(2021)]%
        {2021-LISA}
\bibfield{author}{\bibinfo{person}{Velat Kilic}, \bibinfo{person}{Deepti
  Hegde}, \bibinfo{person}{Vishwanath Sindagi}, \bibinfo{person}{A.~Brinton
  Cooper}, \bibinfo{person}{Mark~A. Foster}, {and} \bibinfo{person}{Vishal~M.
  Patel}.} \bibinfo{year}{2021}\natexlab{}.
\newblock \showarticletitle{Lidar Light Scattering Augmentation {(LISA):}
  Physics-based Simulation of Adverse Weather Conditions for 3D Object
  Detection}.
\newblock \bibinfo{journal}{\emph{CoRR}} \bibinfo{volume}{abs/2107.07004},
  \bibinfo{number}{1} (\bibinfo{year}{2021}), \bibinfo{pages}{1--12}.
\newblock
\urldef\tempurl%
\url{https://arxiv.org/abs/2107.07004}
\showURL{%
\tempurl}


\bibitem[Kim et~al\mbox{.}(2020)]%
        {kim2020reducing}
\bibfield{author}{\bibinfo{person}{Jinhan Kim}, \bibinfo{person}{Jeongil Ju},
  \bibinfo{person}{Robert Feldt}, {and} \bibinfo{person}{Shin Yoo}.}
  \bibinfo{year}{2020}\natexlab{}.
\newblock \showarticletitle{Reducing dnn labelling cost using surprise
  adequacy: An industrial case study for autonomous driving}. In
  \bibinfo{booktitle}{\emph{Proceedings of the 28th ACM Joint Meeting on
  European Software Engineering Conference and Symposium on the Foundations of
  Software Engineering}}. \bibinfo{pages}{1466--1476}.
\newblock
\urldef\tempurl%
\url{https://doi.org/10.1145/3368089.3417065}
\showDOI{\tempurl}


\bibitem[Li et~al\mbox{.}(2023)]%
        {li2023learning}
\bibfield{author}{\bibinfo{person}{Jinlong Li}, \bibinfo{person}{Runsheng Xu},
  \bibinfo{person}{Xinyu Liu}, \bibinfo{person}{Jin Ma},
  \bibinfo{person}{Zicheng Chi}, \bibinfo{person}{Jiaqi Ma}, {and}
  \bibinfo{person}{Hongkai Yu}.} \bibinfo{year}{2023}\natexlab{}.
\newblock \showarticletitle{Learning for vehicle-to-vehicle cooperative
  perception under lossy communication}.
\newblock \bibinfo{journal}{\emph{IEEE Transactions on Intelligent Vehicles}}
  (\bibinfo{year}{2023}).
\newblock
\urldef\tempurl%
\url{https://doi.org/10.1109/TIV.2023.3260040}
\showDOI{\tempurl}


\bibitem[Li et~al\mbox{.}(2018)]%
        {li2018eco}
\bibfield{author}{\bibinfo{person}{Ming Li}, \bibinfo{person}{Xinkai Wu},
  \bibinfo{person}{Xiaozheng He}, \bibinfo{person}{Guizhen Yu}, {and}
  \bibinfo{person}{Yunpeng Wang}.} \bibinfo{year}{2018}\natexlab{}.
\newblock \showarticletitle{An eco-driving system for electric vehicles with
  signal control under V2X environment}.
\newblock \bibinfo{journal}{\emph{Transportation Research Part C: Emerging
  Technologies}}  \bibinfo{volume}{93} (\bibinfo{year}{2018}),
  \bibinfo{pages}{335--350}.
\newblock
\urldef\tempurl%
\url{https://doi.org/10.1016/j.trc.2018.06.002}
\showDOI{\tempurl}


\bibitem[Li et~al\mbox{.}(2022)]%
        {DBLP:journals/corr/abs-2210-05896}
\bibfield{author}{\bibinfo{person}{Shuangzhi Li}, \bibinfo{person}{Zhijie
  Wang}, \bibinfo{person}{Felix Juefei{-}Xu}, \bibinfo{person}{Qing Guo},
  \bibinfo{person}{Xingyu Li}, {and} \bibinfo{person}{Lei Ma}.}
  \bibinfo{year}{2022}\natexlab{}.
\newblock \showarticletitle{Common Corruption Robustness of Point Cloud
  Detectors: Benchmark and Enhancement}.
\newblock \bibinfo{journal}{\emph{CoRR}}  \bibinfo{volume}{abs/2210.05896}
  (\bibinfo{year}{2022}).
\newblock
\urldef\tempurl%
\url{https://doi.org/10.48550/ARXIV.2210.05896}
\showDOI{\tempurl}


\bibitem[Li and Ibanez-Guzman(2020)]%
        {li2020lidar}
\bibfield{author}{\bibinfo{person}{You Li} {and} \bibinfo{person}{Javier
  Ibanez-Guzman}.} \bibinfo{year}{2020}\natexlab{}.
\newblock \showarticletitle{Lidar for autonomous driving: The principles,
  challenges, and trends for automotive lidar and perception systems}.
\newblock \bibinfo{journal}{\emph{IEEE Signal Processing Magazine}}
  \bibinfo{volume}{37}, \bibinfo{number}{4} (\bibinfo{year}{2020}),
  \bibinfo{pages}{50--61}.
\newblock
\urldef\tempurl%
\url{https://doi.org/10.1109/MSP.2020.2973615}
\showDOI{\tempurl}


\bibitem[Liu et~al\mbox{.}(2017)]%
        {liu2017bigroad}
\bibfield{author}{\bibinfo{person}{Luyang Liu}, \bibinfo{person}{Hongyu Li},
  \bibinfo{person}{Jian Liu}, \bibinfo{person}{Cagdas Karatas},
  \bibinfo{person}{Yan Wang}, \bibinfo{person}{Marco Gruteser},
  \bibinfo{person}{Yingying Chen}, {and} \bibinfo{person}{Richard~P Martin}.}
  \bibinfo{year}{2017}\natexlab{}.
\newblock \showarticletitle{Bigroad: Scaling road data acquisition for
  dependable self-driving}. In \bibinfo{booktitle}{\emph{Proceedings of the
  15th Annual International Conference on Mobile Systems, Applications, and
  Services}}. \bibinfo{pages}{371--384}.
\newblock
\urldef\tempurl%
\url{https://doi.org/10.1145/3081333.3081344}
\showDOI{\tempurl}


\bibitem[Liu et~al\mbox{.}(2020)]%
        {liu2020computing}
\bibfield{author}{\bibinfo{person}{Liangkai Liu}, \bibinfo{person}{Sidi Lu},
  \bibinfo{person}{Ren Zhong}, \bibinfo{person}{Baofu Wu},
  \bibinfo{person}{Yongtao Yao}, \bibinfo{person}{Qingyang Zhang}, {and}
  \bibinfo{person}{Weisong Shi}.} \bibinfo{year}{2020}\natexlab{}.
\newblock \showarticletitle{Computing systems for autonomous driving: State of
  the art and challenges}.
\newblock \bibinfo{journal}{\emph{IEEE Internet of Things Journal}}
  \bibinfo{volume}{8}, \bibinfo{number}{8} (\bibinfo{year}{2020}),
  \bibinfo{pages}{6469--6486}.
\newblock
\urldef\tempurl%
\url{https://doi.org/10.1109/JIOT.2020.3043716}
\showDOI{\tempurl}


\bibitem[Liu et~al\mbox{.}(2023)]%
        {liu2023towards}
\bibfield{author}{\bibinfo{person}{Si Liu}, \bibinfo{person}{Chen Gao},
  \bibinfo{person}{Yuan Chen}, \bibinfo{person}{Xingyu Peng},
  \bibinfo{person}{Xianghao Kong}, \bibinfo{person}{Kun Wang},
  \bibinfo{person}{Runsheng Xu}, \bibinfo{person}{Wentao Jiang},
  \bibinfo{person}{Hao Xiang}, \bibinfo{person}{Jiaqi Ma}, {et~al\mbox{.}}}
  \bibinfo{year}{2023}\natexlab{}.
\newblock \showarticletitle{Towards Vehicle-to-everything Autonomous Driving: A
  Survey on Collaborative Perception}.
\newblock \bibinfo{journal}{\emph{arXiv preprint arXiv:2308.16714}}
  (\bibinfo{year}{2023}).
\newblock
\urldef\tempurl%
\url{https://doi.org/10.48550/ARXIV.2308.16714}
\showDOI{\tempurl}


\bibitem[Liu et~al\mbox{.}(2021)]%
        {liu2021automated}
\bibfield{author}{\bibinfo{person}{Wei Liu}, \bibinfo{person}{Xin Xia},
  \bibinfo{person}{Lu Xiong}, \bibinfo{person}{Yishi Lu},
  \bibinfo{person}{Letian Gao}, {and} \bibinfo{person}{Zhuoping Yu}.}
  \bibinfo{year}{2021}\natexlab{}.
\newblock \showarticletitle{Automated vehicle sideslip angle estimation
  considering signal measurement characteristic}.
\newblock \bibinfo{journal}{\emph{IEEE Sensors Journal}} \bibinfo{volume}{21},
  \bibinfo{number}{19} (\bibinfo{year}{2021}), \bibinfo{pages}{21675--21687}.
\newblock
\urldef\tempurl%
\url{https://doi.org/10.1109/JSEN.2021.3059050}
\showDOI{\tempurl}


\bibitem[Ma et~al\mbox{.}(2024)]%
        {ma2024macp}
\bibfield{author}{\bibinfo{person}{Yunsheng Ma}, \bibinfo{person}{Juanwu Lu},
  \bibinfo{person}{Can Cui}, \bibinfo{person}{Sicheng Zhao},
  \bibinfo{person}{Xu Cao}, \bibinfo{person}{Wenqian Ye}, {and}
  \bibinfo{person}{Ziran Wang}.} \bibinfo{year}{2024}\natexlab{}.
\newblock \showarticletitle{MACP: Efficient Model Adaptation for Cooperative
  Perception}. In \bibinfo{booktitle}{\emph{Proceedings of the IEEE/CVF Winter
  Conference on Applications of Computer Vision}}. \bibinfo{pages}{3373--3382}.
\newblock
\urldef\tempurl%
\url{https://doi.org/10.1109/WACV57701.2024.00334}
\showDOI{\tempurl}


\bibitem[McKeeman(1998)]%
        {mckeeman1998differential}
\bibfield{author}{\bibinfo{person}{William~M McKeeman}.}
  \bibinfo{year}{1998}\natexlab{}.
\newblock \showarticletitle{Differential testing for software}.
\newblock \bibinfo{journal}{\emph{Digital Technical Journal}}
  \bibinfo{volume}{10}, \bibinfo{number}{1} (\bibinfo{year}{1998}),
  \bibinfo{pages}{100--107}.
\newblock
\urldef\tempurl%
\url{https://www.hpl.hp.com/hpjournal/dtj/vol10num1/vol10num1art9.pdf}
\showURL{%
\tempurl}


\bibitem[Nasralla et~al\mbox{.}(2014)]%
        {nasralla2014subjective}
\bibfield{author}{\bibinfo{person}{Moustafa~M Nasralla},
  \bibinfo{person}{Chaminda~TER Hewage}, {and} \bibinfo{person}{Maria~G
  Martini}.} \bibinfo{year}{2014}\natexlab{}.
\newblock \showarticletitle{Subjective and objective evaluation and packet loss
  modeling for 3D video transmission over LTE networks}. In
  \bibinfo{booktitle}{\emph{2014 international conference on telecommunications
  and multimedia (temu)}}. IEEE, \bibinfo{pages}{254--259}.
\newblock
\urldef\tempurl%
\url{https://doi.org/10.1109/TEMU.2014.6917770}
\showDOI{\tempurl}


\bibitem[Padilla et~al\mbox{.}(2020)]%
        {padilla2020survey}
\bibfield{author}{\bibinfo{person}{Rafael Padilla}, \bibinfo{person}{Sergio~L
  Netto}, {and} \bibinfo{person}{Eduardo~AB Da~Silva}.}
  \bibinfo{year}{2020}\natexlab{}.
\newblock \showarticletitle{A survey on performance metrics for
  object-detection algorithms}. In \bibinfo{booktitle}{\emph{2020 international
  conference on systems, signals and image processing (IWSSIP)}}. IEEE,
  \bibinfo{pages}{237--242}.
\newblock
\urldef\tempurl%
\url{https://doi.org/10.1109/IWSSIP48289.2020.9145130}
\showDOI{\tempurl}


\bibitem[Rawashdeh and Wang(2018)]%
        {rawashdeh2018collaborative}
\bibfield{author}{\bibinfo{person}{Zaydoun~Yahya Rawashdeh} {and}
  \bibinfo{person}{Zheng Wang}.} \bibinfo{year}{2018}\natexlab{}.
\newblock \showarticletitle{Collaborative automated driving: A machine
  learning-based method to enhance the accuracy of shared information}. In
  \bibinfo{booktitle}{\emph{2018 21st International Conference on Intelligent
  Transportation Systems (ITSC)}}. IEEE, \bibinfo{pages}{3961--3966}.
\newblock
\urldef\tempurl%
\url{https://doi.org/10.1109/ITSC.2018.8569832}
\showDOI{\tempurl}


\bibitem[Segura et~al\mbox{.}(2016)]%
        {segura2016survey}
\bibfield{author}{\bibinfo{person}{Sergio Segura}, \bibinfo{person}{Gordon
  Fraser}, \bibinfo{person}{Ana~B Sanchez}, {and} \bibinfo{person}{Antonio
  Ruiz-Cort{\'e}s}.} \bibinfo{year}{2016}\natexlab{}.
\newblock \showarticletitle{A survey on metamorphic testing}.
\newblock \bibinfo{journal}{\emph{IEEE Transactions on software engineering}}
  \bibinfo{volume}{42}, \bibinfo{number}{9} (\bibinfo{year}{2016}),
  \bibinfo{pages}{805--824}.
\newblock
\urldef\tempurl%
\url{https://doi.org/10.1109/TSE.2016.2532875}
\showDOI{\tempurl}


\bibitem[Stocco et~al\mbox{.}(2023)]%
        {DBLP:journals/ese/StoccoPT23}
\bibfield{author}{\bibinfo{person}{Andrea Stocco}, \bibinfo{person}{Brian
  Pulfer}, {and} \bibinfo{person}{Paolo Tonella}.}
  \bibinfo{year}{2023}\natexlab{}.
\newblock \showarticletitle{Model vs system level testing of autonomous driving
  systems: a replication and extension study}.
\newblock \bibinfo{journal}{\emph{Empir. Softw. Eng.}} \bibinfo{volume}{28},
  \bibinfo{number}{3} (\bibinfo{year}{2023}), \bibinfo{pages}{73}.
\newblock
\urldef\tempurl%
\url{https://doi.org/10.1007/S10664-023-10306-X}
\showDOI{\tempurl}


\bibitem[Wang and Su(2020)]%
        {wang2020metamorphic}
\bibfield{author}{\bibinfo{person}{Shuai Wang} {and} \bibinfo{person}{Zhendong
  Su}.} \bibinfo{year}{2020}\natexlab{}.
\newblock \showarticletitle{Metamorphic Object Insertion for Testing Object
  Detection Systems}. In \bibinfo{booktitle}{\emph{35th {IEEE/ACM}
  International Conference on Automated Software Engineering, {ASE} 2020,
  Melbourne, Australia, September 21-25, 2020}}. \bibinfo{publisher}{{IEEE}},
  \bibinfo{pages}{1053--1065}.
\newblock
\urldef\tempurl%
\url{https://doi.org/10.1145/3324884.3416584}
\showDOI{\tempurl}


\bibitem[Wang et~al\mbox{.}(2020)]%
        {wang2020v2vnet}
\bibfield{author}{\bibinfo{person}{Tsun-Hsuan Wang}, \bibinfo{person}{Sivabalan
  Manivasagam}, \bibinfo{person}{Ming Liang}, \bibinfo{person}{Bin Yang},
  \bibinfo{person}{Wenyuan Zeng}, {and} \bibinfo{person}{Raquel Urtasun}.}
  \bibinfo{year}{2020}\natexlab{}.
\newblock \showarticletitle{V2vnet: Vehicle-to-vehicle communication for joint
  perception and prediction}. In \bibinfo{booktitle}{\emph{Computer
  Vision--ECCV 2020: 16th European Conference, Glasgow, UK, August 23--28,
  2020, Proceedings, Part II 16}}. Springer, \bibinfo{pages}{605--621}.
\newblock
\urldef\tempurl%
\url{https://doi.org/10.1007/978-3-030-58536-5\_36}
\showDOI{\tempurl}


\bibitem[Wen and Jo(2022)]%
        {wen2022deep}
\bibfield{author}{\bibinfo{person}{Li-Hua Wen} {and} \bibinfo{person}{Kang-Hyun
  Jo}.} \bibinfo{year}{2022}\natexlab{}.
\newblock \showarticletitle{Deep learning-based perception systems for
  autonomous driving: A comprehensive survey}.
\newblock \bibinfo{journal}{\emph{Neurocomputing}}  \bibinfo{volume}{489}
  (\bibinfo{year}{2022}), \bibinfo{pages}{255--270}.
\newblock
\urldef\tempurl%
\url{https://doi.org/10.1016/J.NEUCOM.2021.08.155}
\showDOI{\tempurl}


\bibitem[Xie et~al\mbox{.}(2022)]%
        {xie2022towards}
\bibfield{author}{\bibinfo{person}{Xiaoyuan Xie}, \bibinfo{person}{Ying Duan},
  \bibinfo{person}{Songqiang Chen}, {and} \bibinfo{person}{Jifeng Xuan}.}
  \bibinfo{year}{2022}\natexlab{}.
\newblock \showarticletitle{Towards the Robustness of Multiple Object Tracking
  Systems}. In \bibinfo{booktitle}{\emph{2022 IEEE 33rd International Symposium
  on Software Reliability Engineering (ISSRE)}}. IEEE,
  \bibinfo{pages}{402--413}.
\newblock
\urldef\tempurl%
\url{https://doi.org/10.1109/ISSRE55969.2022.00046}
\showDOI{\tempurl}


\bibitem[Xie et~al\mbox{.}(2011a)]%
        {xie2011testing}
\bibfield{author}{\bibinfo{person}{Xiaoyuan Xie}, \bibinfo{person}{Joshua
  Wing~Kei Ho}, \bibinfo{person}{Christian Murphy}, \bibinfo{person}{Gail~E.
  Kaiser}, \bibinfo{person}{Baowen Xu}, {and} \bibinfo{person}{Tsong~Yueh
  Chen}.} \bibinfo{year}{2011}\natexlab{a}.
\newblock \showarticletitle{Testing and validating machine learning classifiers
  by metamorphic testing}.
\newblock \bibinfo{journal}{\emph{J. Syst. Softw.}} \bibinfo{volume}{84},
  \bibinfo{number}{4} (\bibinfo{year}{2011}), \bibinfo{pages}{544--558}.
\newblock
\urldef\tempurl%
\url{https://doi.org/10.1016/j.jss.2010.11.920}
\showDOI{\tempurl}


\bibitem[Xie et~al\mbox{.}(2011b)]%
        {xie2011spectrum}
\bibfield{author}{\bibinfo{person}{Xiaoyuan Xie}, \bibinfo{person}{W.~Eric
  Wong}, \bibinfo{person}{Tsong~Yueh Chen}, {and} \bibinfo{person}{Baowen Xu}.}
  \bibinfo{year}{2011}\natexlab{b}.
\newblock \showarticletitle{Spectrum-Based Fault Localization: Testing Oracles
  are No Longer Mandatory}. In \bibinfo{booktitle}{\emph{Proceedings of the
  11th International Conference on Quality Software, {QSIC} 2011, Madrid,
  Spain, July 13-14, 2011}}, \bibfield{editor}{\bibinfo{person}{Manuel
  N{\'{u}}{\~{n}}ez}, \bibinfo{person}{Robert~M. Hierons}, {and}
  \bibinfo{person}{Mercedes~G. Merayo}} (Eds.). \bibinfo{publisher}{{IEEE}
  Computer Society}, \bibinfo{pages}{1--10}.
\newblock
\urldef\tempurl%
\url{https://doi.org/10.1109/QSIC.2011.20}
\showDOI{\tempurl}


\bibitem[Xie et~al\mbox{.}(2013)]%
        {xie2013metamorphic}
\bibfield{author}{\bibinfo{person}{Xiaoyuan Xie}, \bibinfo{person}{W.~Eric
  Wong}, \bibinfo{person}{Tsong~Yueh Chen}, {and} \bibinfo{person}{Baowen Xu}.}
  \bibinfo{year}{2013}\natexlab{}.
\newblock \showarticletitle{Metamorphic slice: An application in spectrum-based
  fault localization}.
\newblock \bibinfo{journal}{\emph{Inf. Softw. Technol.}} \bibinfo{volume}{55},
  \bibinfo{number}{5} (\bibinfo{year}{2013}), \bibinfo{pages}{866--879}.
\newblock
\urldef\tempurl%
\url{https://doi.org/10.1016/j.infsof.2012.08.008}
\showDOI{\tempurl}


\bibitem[Xu et~al\mbox{.}(2023)]%
        {xu2023v2v4real}
\bibfield{author}{\bibinfo{person}{Runsheng Xu}, \bibinfo{person}{Xin Xia},
  \bibinfo{person}{Jinlong Li}, \bibinfo{person}{Hanzhao Li},
  \bibinfo{person}{Shuo Zhang}, \bibinfo{person}{Zhengzhong Tu},
  \bibinfo{person}{Zonglin Meng}, \bibinfo{person}{Hao Xiang},
  \bibinfo{person}{Xiaoyu Dong}, \bibinfo{person}{Rui Song}, {et~al\mbox{.}}}
  \bibinfo{year}{2023}\natexlab{}.
\newblock \showarticletitle{V2v4real: A real-world large-scale dataset for
  vehicle-to-vehicle cooperative perception}. In
  \bibinfo{booktitle}{\emph{Proceedings of the IEEE/CVF Conference on Computer
  Vision and Pattern Recognition}}. \bibinfo{pages}{13712--13722}.
\newblock
\urldef\tempurl%
\url{https://doi.org/10.1109/CVPR52729.2023.01318}
\showDOI{\tempurl}


\bibitem[Xu et~al\mbox{.}(2022a)]%
        {xu2022v2x}
\bibfield{author}{\bibinfo{person}{Runsheng Xu}, \bibinfo{person}{Hao Xiang},
  \bibinfo{person}{Zhengzhong Tu}, \bibinfo{person}{Xin Xia},
  \bibinfo{person}{Ming-Hsuan Yang}, {and} \bibinfo{person}{Jiaqi Ma}.}
  \bibinfo{year}{2022}\natexlab{a}.
\newblock \showarticletitle{V2x-vit: Vehicle-to-everything cooperative
  perception with vision transformer}. In \bibinfo{booktitle}{\emph{European
  conference on computer vision}}. Springer, \bibinfo{pages}{107--124}.
\newblock
\urldef\tempurl%
\url{https://doi.org/10.1007/978-3-031-19842-7\_7}
\showDOI{\tempurl}


\bibitem[Xu et~al\mbox{.}(2022b)]%
        {xu2022opv2v}
\bibfield{author}{\bibinfo{person}{Runsheng Xu}, \bibinfo{person}{Hao Xiang},
  \bibinfo{person}{Xin Xia}, \bibinfo{person}{Xu Han}, \bibinfo{person}{Jinlong
  Li}, {and} \bibinfo{person}{Jiaqi Ma}.} \bibinfo{year}{2022}\natexlab{b}.
\newblock \showarticletitle{Opv2v: An open benchmark dataset and fusion
  pipeline for perception with vehicle-to-vehicle communication}. In
  \bibinfo{booktitle}{\emph{2022 International Conference on Robotics and
  Automation (ICRA)}}. IEEE, \bibinfo{pages}{2583--2589}.
\newblock
\urldef\tempurl%
\url{https://doi.org/10.1109/ICRA46639.2022.9812038}
\showDOI{\tempurl}


\bibitem[Yu et~al\mbox{.}(2021)]%
        {yu2021edge}
\bibfield{author}{\bibinfo{person}{Ruozhou Yu}, \bibinfo{person}{Dejun Yang},
  {and} \bibinfo{person}{Hao Zhang}.} \bibinfo{year}{2021}\natexlab{}.
\newblock \showarticletitle{Edge-assisted collaborative perception in
  autonomous driving: A reflection on communication design}. In
  \bibinfo{booktitle}{\emph{2021 IEEE/ACM Symposium on Edge Computing (SEC)}}.
  IEEE, \bibinfo{pages}{371--375}.
\newblock


\bibitem[Yurtsever et~al\mbox{.}(2020)]%
        {yurtsever2020survey}
\bibfield{author}{\bibinfo{person}{Ekim Yurtsever}, \bibinfo{person}{Jacob
  Lambert}, \bibinfo{person}{Alexander Carballo}, {and} \bibinfo{person}{Kazuya
  Takeda}.} \bibinfo{year}{2020}\natexlab{}.
\newblock \showarticletitle{A survey of autonomous driving: Common practices
  and emerging technologies}.
\newblock \bibinfo{journal}{\emph{IEEE access}}  \bibinfo{volume}{8}
  (\bibinfo{year}{2020}), \bibinfo{pages}{58443--58469}.
\newblock
\urldef\tempurl%
\url{https://doi.org/10.1109/ACCESS.2020.2983149}
\showDOI{\tempurl}


\bibitem[Zhang et~al\mbox{.}(2024)]%
        {DBLP:journals/tosem/ZhangZFLSHW24}
\bibfield{author}{\bibinfo{person}{Quanjun Zhang}, \bibinfo{person}{Juan Zhai},
  \bibinfo{person}{Chunrong Fang}, \bibinfo{person}{Jiawei Liu},
  \bibinfo{person}{Weisong Sun}, \bibinfo{person}{Haichuan Hu}, {and}
  \bibinfo{person}{Qingyu Wang}.} \bibinfo{year}{2024}\natexlab{}.
\newblock \showarticletitle{Machine Translation Testing via Syntactic Tree
  Pruning}.
\newblock \bibinfo{journal}{\emph{{ACM} Trans. Softw. Eng. Methodol.}}
  \bibinfo{volume}{33}, \bibinfo{number}{5} (\bibinfo{year}{2024}),
  \bibinfo{pages}{125:1--125:39}.
\newblock
\urldef\tempurl%
\url{https://doi.org/10.1145/3640329}
\showDOI{\tempurl}


\end{thebibliography}

\end{document}